\documentclass[a4paper,fleqn,usenatbib]{mnras}
\usepackage{newtxtext,newtxmath}
\usepackage[T1]{fontenc}
\usepackage{ae,aecompl}
\usepackage{multicol}
\usepackage{graphicx}	
\usepackage{amsmath}	
\usepackage{cleveref}
\usepackage{pdflscape}

\newcommand{\kms}{{\mathrm{km~s^{-1}}}}

\title[Mode identification and seismic study of $\delta$ Scuti]{Mode identification and seismic study of $\delta$ Scuti, the prototype
of a class of pulsating stars}

\author[Daszy\'nska-Daszkiewicz et al.]{ J. Daszy\'nska-Daszkiewicz$^{1}$\thanks{E-mail:daszynska@astro.uni.wroc.pl},
A. A. Pamyatnykh$^{2}$, P. Walczak$^{1}$,  G. Handler$^{2}$,
\newauthor A. Pigulski$^{1}$, W. Szewczuk$^{1}$,\\
$^{1}$Instytut Astronomiczny, Uniwersytet Wroc{\l}awski, Kopernika 11, 51-622 Wroc{\l}aw, Poland\\
$^{2}$Nicolaus Copernicus Astronomical Center, Polish Academy of Sciences, Bartycka 18, 00-716 Warsaw, Poland\\
}

\date{Accepted XXX. Received YYY; in original form ZZZ}

\pubyear{2021}

\begin{document}
\label{firstpage}
\pagerange{\pageref{firstpage}--\pageref{lastpage}}
\maketitle

\begin{abstract}
We present a seismic study of $\delta$ Scuti based on a mode identification from multicoulor photometry.
The dominant frequency can be associated only with a radial mode and the second frequency is, most probably, a dipole mode.
The other six frequencies have more ambiguous identifications.
The photometric mode identification provided also some constraints on the atmospheric metallicity [m/H]$\approx$+0.5
and microturbulent velocity $\xi_t\approx 4~\kms$.\\
For models reproducing the dominant frequency, we show that only the fundamental mode is possible and the first overtone is excluded.
However, the location of $\delta$ Scuti near the terminal age main sequence requires the consideration of three stages of stellar evolution.
For the star to be on the main sequence, it is necessary to include overshooting from the convective core with a parameter of at least $\alpha_{\rm ov}=0.25$ at the metallicity greater than $Z=0.019$.
It turned out that the value of the relative amplitude of the bolometric flux variations
(the nonadiabatic parameter $f$) is mainly determined by the position of the star in the HR diagram, i.e., by its effective temperature
and luminosity, whereas the effect of the evolutionary stage is minor.
On the other hand, the convective efficiency in the subphotospheric layers
has a dominant effect on the value of the parameter $f$. 
Comparing the theoretical and empirical values of $f$ for the radial dominant mode, we obtain constraints on the mixing length
parameter $\alpha_{\rm MLT}$ which is less than about 1.0, independently of the adopted opacity data and chemical mixture.
This value of $\alpha_{\rm MLT}$ is substantially smaller than for a calibrated solar model
indicating rather low to moderately efficient convection in the envelope of $\delta$ Scuti.
\end{abstract}



\section{Introduction}

$\delta$ Scuti stars are classical pulsators with masses in the range of about 1.6 - 2.6 $\mathrm{M}_\odot$.
Most of them are in the main-sequence phase of evolution but there are $\delta$ Sct stars that have already
completed central hydrogen burning \citep[e.g.,][]{BregerPam1998, Bowman2016} or $\delta$~Sct stars in the pre-main sequence contraction stage \citep[e.g.,][]{Marconi1998, Zwintz2014}.
The pulsational driving mechanism of these variables has been understood a long time ago \citep{Chevalier1971}.
This instability is due to the opacity mechanism operating in the second helium ionization zone but
there may also be a small contribution to driving from the hydrogen ionization region \citep{Pamyatnykh1999}.
Radial pulsations and non-radial pulsations in pressure (p) and gravity (g) modes can be excited.

In most $\delta$ Scuti stars multiperiodicity is observed which potentially permits the derivation of
stringent seismic constraints on their internal structure and evolution.
Before the era of space-based observations of pulsating stars, the $\delta$ Sct star with the most detected pulsation frequencies was FG Vir \citep{Breger2005}, with almost 70 independent oscillation modes. The richness of the oscillations of the $\delta$ Sct variables was revealed
by the satellite missions CoRoT and Kepler. For some objects hundreds of frequencies were extracted from the light curves. For example, the CoRoT observations of HD 50844 revealed hundreds of peaks in the frequency range 0-30 d$^{-1}$ \citep{Poretti2009}.
In particular, the detection of high-order g modes in most $\delta$ Scuti stars observed by Kepler \citep{Grigahcene2010}
required a new approach to pulsational modelling because such modes
are stable in all stellar models computed with the standard opacity data \citep{Balona2015}. Using Gaia DR2 parallaxes and Kepler data for a sample of over 15000 A- and F-type stars, \citet{Murphy2019} classified them into $\delta$ Scuti
and non-$\delta$ Scuti stars finding that many stars in the classical instability strip do not pulsate. They defined a new empirical instability strip in the Hertzsprung-Russell diagram that is systematically hotter than the theoretical strip.
\citet{Bowman2018} also found that a significant fraction of main-sequence $\delta$ Scuti stars is located outside of the classical instability strip in the HR diagram. To explain both low- and high-order pressure modes in $\delta$ Scuti stars, the mass dependence of the convective mixing length parameter may be important. A list of $\delta$ Scuti variable star candidates from the  K2 mission as targets for asteroseismic studies was prepared by \citet{Guzik2019}. \citet{Antoci2019} classified and studied pulsation properties of a sample of 117 $\delta$~Scuti and $\gamma$~Doradus stars observed by the {\it TESS} mission. In particular, they stressed the important role of mixing processes in the outer stellar layers on pulsation driving.

Detailed seismic modelling of $\delta$ Scuti stars, in particular those observed from space,
is hampered by a lack of mode identification. Except for a few cases of stars with some
regular patterns in oscillation spectra, unambiguous identification of modes comes only
from ground-based photometric and/or spectroscopic data. However, not all frequency peaks observed from space
are detectable from the ground.
Moreover, in $\delta$ Sct star models, the photometric amplitudes and phases are very sensitive
to the efficiency of convective transport in the outer layers. This sensitivity occurs via the nonadiabatic parameter $f$ that
gives the relative amplitude of the radiative flux perturbation at the photospheric level.
To by-pass this effect, \citet{JDD2003} invented the method
of simultaneous determination of the mode degree $\ell$ and the parameter $f$ from multi-colour data.
Thinking the other way, one can expect constraints on the efficiency of envelope convection from a comparison of the theoretical
and empirical values of $f$. Such constraints have been derived for several $\delta$ Scuti stars \citep{JDD2003, JDD2005, JDD2007},
and, recently, for SX~Phoenicis, the prototype of Population II counterparts of $\delta$~Sct stars \citep{JDD2020}.
The general conclusion was that convective transport in the envelope of SX~Phe is poor to moderately efficient.

Here, we present the first mode identification based on multi-colour photometry for the prototype star $\delta$ Scuti.
In Sect.\,2, we give  brief description and literature survey for the star. Sect.\,3 contains
details of observations and frequency analysis. In Sect.\,4, we present the mode identification
based on the photometric amplitudes and phases.
Pulsational models reproducing the dominant frequency are discussed in Sect.\,5.
Complex seismic modelling and discussions of various effects on the parameter $f$ are given in Sect.\,6.
The summary and some conclusions are in the last section.

\section{The prototype $\delta$ Scuti}

The class prototype $\delta$ Scuti (HD 172748) has a brightness $V=4.71$\,mag and a
spectral type F2II-III \citep{Gray2001}.
Its variable radial velocity was detected over 120 years ago by \citet{Campbell1900}.
$\delta$ Scuti is a quite nearby object with a Gaia EDR3 parallax
$\pi=16.1899(1157)$ mas translating into a distance $d=61.8(4)$\,pc
\citep{GaiaDR3}.
There are several determinations of its effective temperature in the literature ranging
from about 6770\,K \citep{Erspamer2003} to about 7260\,K \citep[e.g.,][]{Balona1994}. The most recent determinations
amount to $T_{\rm eff}\approx7000$\,K \citep[e.g.,][]{Yushchenko2005, Holmberg2009, Boeche2016}.
The surface gravity of $\delta$ Scuti is $\log g\approx 3.5$ \citep{Erspamer2003, Yushchenko2005}.
In our study, we allow for a whole range of $T_{\rm eff}$,  i.e., 6770 - 7260\,K, with a logarithmic value $\log T_{\rm eff}=3.845(15)$.
\begin{figure}
\includegraphics[width=\columnwidth,clip]{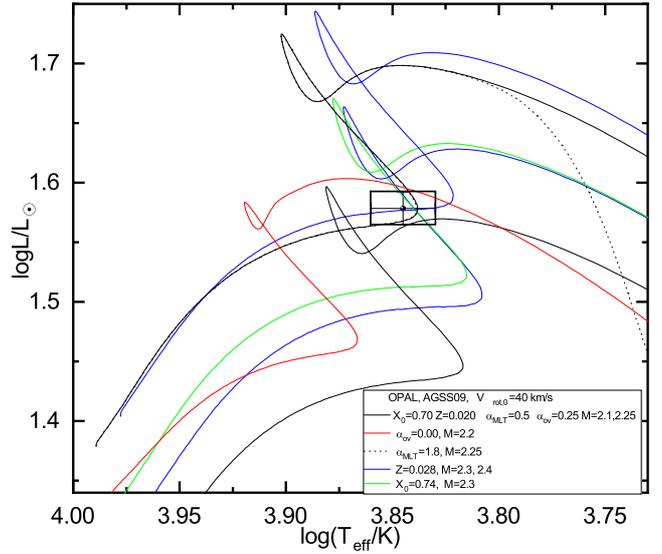}
\caption{The HR diagram with the position of $\delta$ Scuti as determined from the whole observed range of the effective temperature
and luminosity. The evolutionary tracks were computed for OPAL opacities with the AGSS09
solar mixture at various combinations of the initial hydrogen abundance $X_0$, metallicity $Z$, overshooting from the convective core
$\alpha_{\rm ov}$ and the mixing length parameter in the envelope $\alpha_{\rm MLT}$, as indicated in the legend.}
\label{fig1}
\end{figure}

According to \citet{Yushchenko2005}, the metallicity of $\delta$ Sct is above the solar value
and its abundance pattern is similar to those of Am-Fm type stars.
\citet{Erspamer2003} classified $\delta$ Sct among metallic giants, which they defined as evolved giants
with a surplus of such elements as: Al, Ca, Ti, Cr, Mn, Fe, Ni and Ba.
In the case of $\delta$ Scuti, all of these elements, except Ca, show an excess comparing to the solar values.
The estimated values of metallicity in the literature are  $\mbox{[Fe/H]}=0.41$,  by \citet{Holmberg2009}
from the calibration of $uvby\beta$ photometry, and $\mbox{[m/H]}\in(0.18, 0.38)$
by \citet{Boeche2016} from spectroscopic analysis adopting solar abundances of \citet{GrevSau1998}. The microturbulent velocity of 2.8\,km\,s$^{-1}$ is given by \citet{Erspamer2003} and 3.8\,km\,s$^{-1}$ by \citet{Yushchenko2005}.
If the inclination of the rotation axis is not low, $\delta$ Sct is a rather slow rotator with a projected rotational velocity
$V_{\rm rot}\sin i=25~\kms$ \citep{Erspamer2003, Schroeder2009}.

Using the Gaia EDR3 parallax and the bolometric correction from Kurucz models with a proper zero point, i.e., $M_{\rm bol,\sun}=4.62$
\citep{Torres2010}, we derived the luminosity considering three values of the atmospheric metallicity $\mbox{[m/H]}=0.2,~0.3,~0.5$
and the microturbulent velocity $\xi_t=2~\kms$.
Unfortunately, there are no Kurucz models for $\mbox{[m/H]}=0.4$ and for higher values of $\xi_t$ at the metallicity $\mbox{[m/H]}\ge 0.2$.
We obtained the following values:  $\log L/L_{\sun}= 1.584(12)$ for $\mbox{[m/H]}=0.2$,
$\log L/L_{\sun}= 1.580(12)$ for $\mbox{[m/H]}=0.3$ and $\log L/L_{\sun}= 1.574(13)$ for $\mbox{[m/H]}=0.5$.
For the further studies, we adopted the total range, that is $\log L/L_{\sun}= 1.578(17)$.

To convert the metallicity $\mbox{[m/H]}$ into the metal mass fraction $Z$, one can use a simple approximate formula
$$\mbox{[m/H]}=\log(Z/Z_{\sun}) - \log(X/X_{\sun}).$$
The value of $\mbox{[m/H]}\in(0.18, 0.38)$, as determined by \citet{Boeche2016} translates into $Z=0.031\pm0.07$ at $X=0.70$,
taking the values of $X_{\sun}=0.7345$ and $Z_{\sun}=0.0169$ from \citet{GrevSau1998}.
However, one has to be aware that the chemical anomalies are observed in the atmosphere of $\delta$ Scuti and the bulk metallicity of the star
can be much lower or even around the solar value. Therefore, in our analysis we will consider also lower values of $Z$ than those obtained above.

The position of $\delta$ Scuti in the HR diagram is shown in Fig.\,1.
The evolutionary tracks were computed using the Warsaw-New Jersey code \citep[e.g.,][]{Pamyatnykh1998, Pamyatnykh1999}
assuming the OPAL opacity tables \citep{Iglesias1996} and the opacities of \citet{Ferguson2005} for temperatures $\log T<3.95$.
In all calculations the solar chemical mixture was adopted from \citet{Asplund2009} (AGSS09) and the OPAL2005 equation of state was used \citep{Rogers1996, Rogers2002}. The Warsaw-New Jersey code takes into account the mean effect of the centrifugal force,
assuming solid-body rotation and constant global angular momentum during evolution.
The treatment of convection relies on the standard mixing-length theory.

We depicted the tracks for masses in the range [2.1,\,2.4] $\mathrm{M}_\odot$ considering different values of the initial
hydrogen abundance $X_0=0.70,~0.74$, metallicity $Z=0.020,~0.028$, overshooting from the convective core $\alpha_{\rm ov}=0.0,~0.25$
and the mixing length parameter for envelope convection $\alpha_{\rm MLT}=0.5,~1.8$, to demonstrate the effects of these parameters.
In all cases the initial rotational velocity of $V_{\rm rot}=40~\kms$ was assumed to reach the value of about $25-30~\kms$
within the error box of the position of $\delta$ Scuti in the HR diagram.

As one can see, the star is located around the Terminal Age Main Sequence (TAMS) which means that
three evolutionary phases have to be considered: main sequence (MS), overall contraction (OC) and post-MS.
To consider $\delta$ Scuti models in the MS phase, convective overshooting of at least $\alpha_{\rm ov}=0.25$ is indispensable.

Despite being the prototype of the class, $\delta$ Scuti has so far been insufficiently observed and studied.
Up to now six frequencies were detected in its light curve in the range 4.6 -- 8.6 d$^{-1}$
by \citet{Templeton1997} who carried out a five-year Str\"omgren $y$ photometric campaign.
Moreover, \citet{Templeton1997} constructed a grid of evolutionary and nonadiabatic pulsational models. For their best-fit models, they associated the highest amplitude frequency $\nu_1=5.16$ d$^{-1}$ with
the radial fundamental mode in contrast to the results of \citet{Balona1981} and \citet{Cugier1993} that
the strongest mode is the radial first overtone. Furthermore, \citet{Templeton1997} obtained that
all modes with frequencies in the observed range are unstable.
However, without observations in more passbands an independent mode identification was impossible.
\citet{Cugier1993} used the phase shifts between radial
velocity data and light curves obtained from IUE and infrared data in order to identify the modes.
Their preliminary mode identification for the second frequency $\nu_2=5.35$ d$^{-1}$ was a $\ell=2,~p_2$ mode.
Apart from the works of \citet{Templeton1997} and \citet{Cugier1993}, there were no more studies of the properties
of the pulsational modes of $\delta$ Scuti.

\section{Space and ground-based photometry}
As mentioned above, \cite{Templeton1997} carried out a photometric campaign observing $\delta$~Scuti in the years 1983\,--\,1988 in the Str\"omgren $y$ filter. They discovered eight terms in their photometry, including six independent frequencies, one harmonic of the strongest frequency and one combination frequency. Their data suffered from strong daily and yearly aliasing, leaving some ambiguity in the frequencies of all but the two strongest peaks. In the present paper, we analyse the three photometric data sets. Firstly, we use Solar Mass Ejection Imager (SMEI) space wide-band photometry to derive precise frequencies of pulsations.
Then, the Str\"omgren $uvy$ ground-based data gathered by us at Fairborn Observatory are analysed to obtain multi-band amplitudes and phases of the observed frequencies. Finally, we re-analyse $y$-filter photometry of \cite{Templeton1997}.

\subsection{SMEI photometry}\label{sect:smei}
The SMEI experiment placed aboard the Coriolis spacecraft \citep{Eyles2003,Jackson2004} measured sunlight scattered by free electrons in the solar wind scanning the whole sky with three wide-field cameras. A by-product of the mission was time-series photometry of over 5500 bright stars obtained in the years 2003\,--\,2010. The SMEI photometry is affected by long-term calibration effects, especially a repeatable variability with a period of one year. The raw SMEI photometry of $\delta$~Sct, downloaded from the University of California San Diego (UCSD) web page\footnote{http://smei.ucsd.edu/new\_smei/index.html}, was corrected for the one-year variability by subtracting an interpolated mean light curve. The mean light curve was obtained by folding the raw data with the period of one year, calculating median values in 200 intervals in phase, and then interpolating between them.
The interpolated curve was then subtracted from the data.
The subsequent procedure was aimed at further removal of instrumental effects and consisted of the following steps: (i) Identification of the strongest frequencies in the frequency spectrum. (ii) Fitting a model consisting of a sum of sinusoids with the detected frequencies to the data and subtraction of this model from the data. The next three steps were carried out using residuals from this fit. (iii) Identification and rejection of outliers by means of the Generalized Extreme Studentized Deviate algorithm \citep{Rosner1983}. (iv) Calculation of individual uncertainties using the scatter of the neighbouring data. The data with uncertainties higher than a subjectively chosen threshold were then removed from the dataset. (v) Removal of low-frequency signals, primarily of instrumental origin, by the calculation of means in time intervals, interpolating between them and subtracting. Steps (iii) to (v) were iterated several times, yielding 22795 data points that were used in the final analysis. The final uncertainties ranged between 5 and 25~mmag.
\begin{figure}
	\includegraphics[width=\columnwidth]{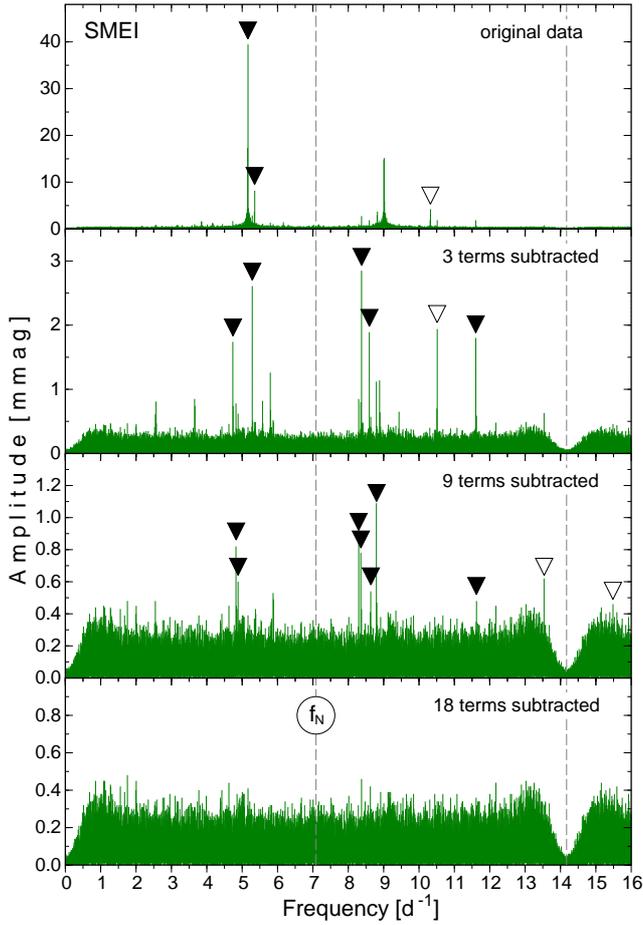}
	\caption{Fourier frequency spectra from the SMEI data. Going from top to bottom, the panels show four steps of prewhitening. The independent frequencies are shown
		with filled triangles, whereas the harmonics of $\nu_1$ and combination frequencies with open triangles. Vertical dashed lines mark the satellite orbital frequency
		$f_{\rm orb}= 14.172$\,d$^{-1}$ and related Nyquist frequency $f_{\rm N}=f_{\rm orb}/2=7.086$\,d$^{-1}$.}
	\label{fig:smei-fs}
\end{figure}

Analysis of the SMEI time-series photometry corrected for instrumental effects resulted in detection of 18 terms. We applied a standard prewhitening procedure, in which frequency spectra were calculated at each step with the use of a Discrete Fourier Transform. At each step of the procedure, all previously detected terms were added to the model consisting of a sum of sinusoidal terms. Their amplitudes and phases were derived by fitting the model to the original data by means of the least squares method. In the next step, the residuals from the fit were used to calculate the next frequency spectrum. Frequency spectra at several steps of prewhitening are shown in Fig.\,\ref{fig:smei-fs}. Of all detected terms, 14 represent independent frequencies,
two are harmonics of the highest-amplitude frequency, $\nu_1$, and the remaining two are combination frequencies. The frequencies and amplitudes of all terms are listed in Table \ref{tab:freqs}.

As can be seen in Fig.\,\ref{fig:smei-fs} and Table \ref{tab:freqs}, the independent frequencies extracted from the SMEI data occur in three relatively narrow ranges;
between 4.74 and 5.35\,d$^{-1}$ (6 peaks), between 8.29 and 8.80\,d$^{-1}$ (6 peaks), and near 11.6\,d$^{-1}$ (2 peaks). A sudden drop of signal below $\sim$0.7\,d$^{-1}$
and its replicas around the satellite's orbital frequency $f_{\rm orb}= 14.172$\,d$^{-1}$ (Fig.\,\ref{fig:smei-fs}) are the result of the strong detrending applied to remove the instrumental effects.
If some low-amplitude peaks would be present at these frequencies, they would escape detection.

\subsection{APT photometry}\label{sect:apt}
The star was also observed photometrically with the 0.75-m T6 automatic telescope at Fairborn Observatory (Arizona, USA) in three passbands, $u$, $v$, and $y$, of the Str\"omgren photometric system. Two comparison stars, HD\,173093  ($V=6.30$~mag) and HD\,174464 ($V=5.84$~mag), were used. No intrinsic variability of any of those two stars was detected within a limit of 1.2\,mmag in Fourier space in the range of the pulsation frequencies of the prime target. Observations started on May 12, 2018 and were gathered during 36 nights, covering the interval of about 46 days. In Fig.\,3, we plot examples of light curves in all three photometric bands.
\begin{figure*}
	\includegraphics[width=175mm,height=12cm,clip]{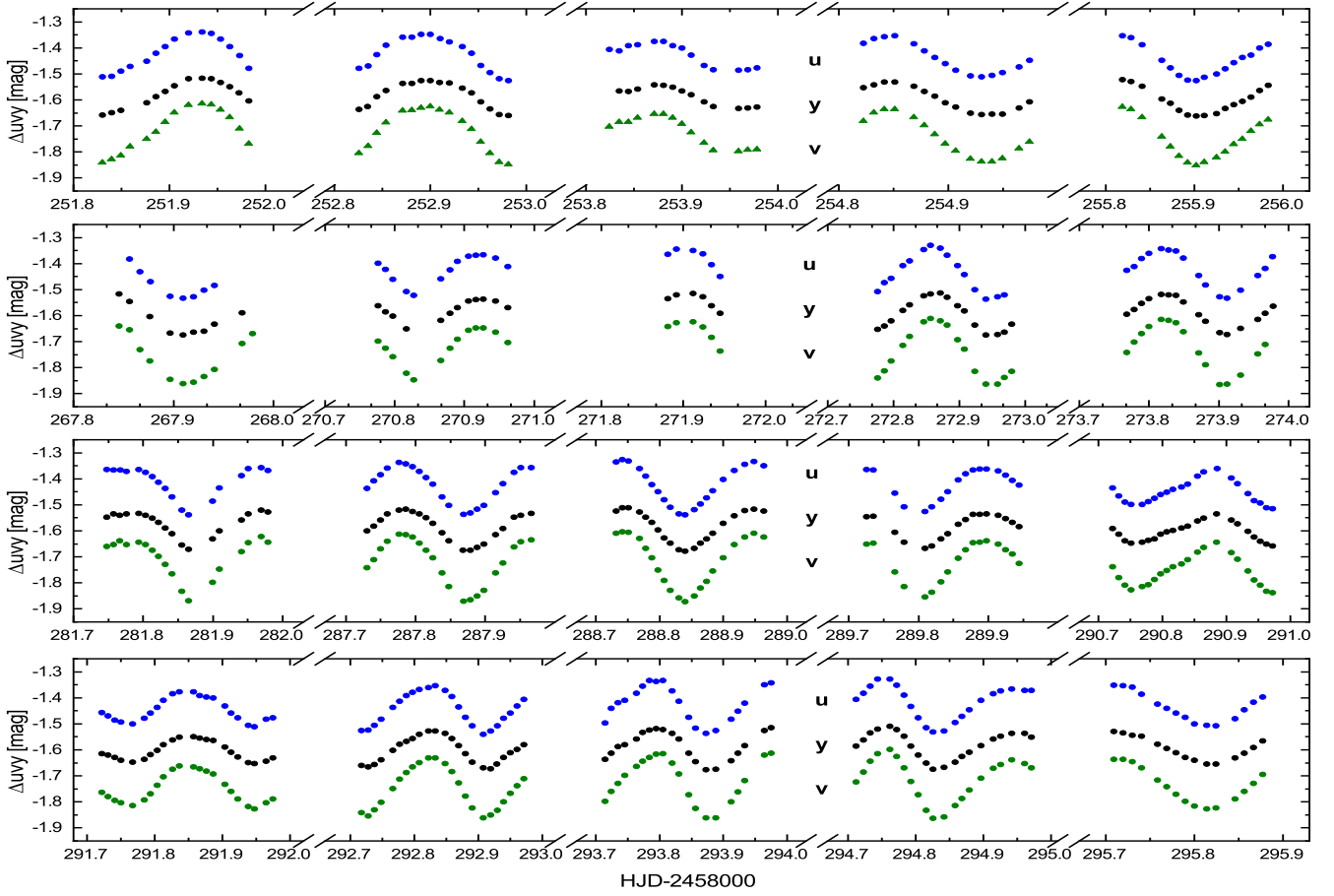}
	\caption{Examples of the APT observations of $\delta$ Scuti in the Str\"omgren $uvy$ passbands.}
	\label{fig3}
\end{figure*}

\begin{figure}
\includegraphics[width=\columnwidth]{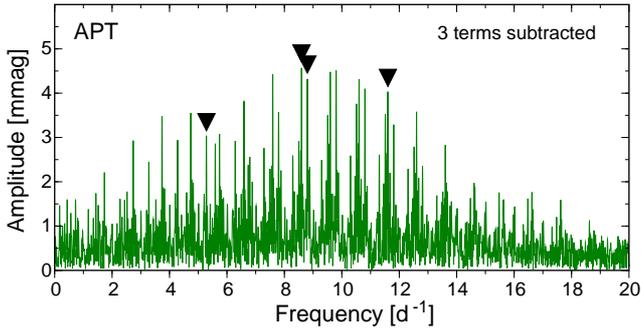}
\caption{Fourier frequency spectrum of the $y$-filter APT data after subtracting the three strongest terms, $\nu_1$, $\nu_2$, and 2$\nu_1$. The spectrum illustrates the problem of
strong daily aliasing. The true frequencies of four fainter terms, $\nu_4$, $\nu_5$, $\nu_6$, and $\nu_8$, are indicated by inverted triangles.}
\label{fig:apt-fs}
\end{figure}

The analysis of the APT data was carried out independently for the three passbands. Since single-site APT data suffer from strong daily aliasing, during the consecutive steps of prewhitening the maxima with frequency equivalents obtained from the SMEI data were considered to be real. These were not always the highest peaks. The aliasing problem  in the APT data is illustrated in Fig.\,\ref{fig:apt-fs}, in which we show the frequency spectrum of the $y$-filter data after subtraction of the three strongest terms, $\nu_1$, $\nu_2$, and 2$\nu_1$. The problem is compounded
 by the fact that frequencies of two terms, $\nu_5$ and $\nu_6$, are separated by almost exactly 3\,d$^{-1}$, so that their daily aliases coincide. We detected the same 10 terms in all three passbands of the APT data; their amplitudes are given in Table~\ref{tab:freqs}. The detected terms were the strongest in the SMEI data, except for $\nu_3$ that was not detected in the APT data. Such a huge amplitude drop, below the detection level, may be due to non-linear interaction of various pulsational modes \citep{Moskalik1985}.

\begin{landscape}
	\begin{table}
		\centering
		\tabcolsep=4pt
		\caption{Sinusoidal terms detected in the photometric data of $\delta$~Scuti. All frequencies but $\nu_{15}$ were detected in the SMEI data, see text for explanation. Starting from the third column, the values of signal-to-noise ratio (S/N), semi-amplitude, $A$, and phase, $\phi$, are given. The noise level in the frequency spectrum of residuals was calculated as a mean in the range 0\,--\,25\,d$^{-1}$ for the APT and MTy data and 1\,--\,13\,d$^{-1}$ for the SMEI data.
			Phases are given for the following epochs: HJD\,2454000 for the SMEI data, HJD\,2458270 for the APT data, and HJD\,2446493.5 for the data of \protect\cite{Templeton1997}. Subscripts identify data set: SMEI data are labeled with `SMEI', APT $uvy$ data with `$u$', `$v$', and `$y$', respectively, and $y$-filter data of \protect\cite{Templeton1997}, with `$y$,T'. The final three rows of the table contain standard deviation of residuals, SDR, detection threshold, DT, defined as four times the noise level in the frequency spectrum of the residuals, and the number of observations used, $N_{\rm obs}$. Numbers in parentheses represent uncertainties of the preceding values, with the leading zeroes omitted.}
		\begin{tabular}{cr@{.}lrrr@{.}lrrr@{.}lrrr@{.}lrrr@{.}lrrr@{.}l}
			\hline
			\multicolumn{1}{c}{ID} & \multicolumn{2}{c}{Frequency}  & \multicolumn{1}{c}{S/N} & \multicolumn{1}{c}{$A_{\rm SMEI}$} & \multicolumn{2}{c}{$\phi_{\rm SMEI}$} &\multicolumn{1}{c}{S/N} & \multicolumn{1}{c}{$A_u$} & \multicolumn{2}{c}{$\phi_u$} & \multicolumn{1}{c}{S/N} & \multicolumn{1}{c}{$A_v$} & \multicolumn{2}{c}{$\phi_v$} & \multicolumn{1}{c}{S/N} & \multicolumn{1}{c}{$A_y$} & \multicolumn{2}{c}{$\phi_y$} & \multicolumn{1}{c}{S/N} & \multicolumn{1}{c}{$A_{y,{\rm T}}$} & \multicolumn{2}{c}{$\phi_{y,{\rm T}}$}\\
			& \multicolumn{2}{c}{[d$^{-1}$]} & \multicolumn{1}{c}{(SMEI)} & \multicolumn{1}{c}{[mmag]} & \multicolumn{2}{c}{[rad]} & \multicolumn{1}{c}{($u$)} & \multicolumn{1}{c}{[mmag]} & \multicolumn{2}{c}{[rad]} & \multicolumn{1}{c}{($v$)} & \multicolumn{1}{c}{[mmag]} & \multicolumn{2}{c}{[rad]} & \multicolumn{1}{c}{($y$)} & \multicolumn{1}{c}{[mmag]} & \multicolumn{2}{c}{[rad]} & \multicolumn{1}{c}{($y$,T)} & \multicolumn{1}{c}{[mmag]} & \multicolumn{2}{c}{[rad]}\\
			\hline
			$\nu_1$ & 5&1607680(6)  & 357.8 & 39.64(12) & 0&1109(29) &  153.7 & 80.99(40) & 3&238(5) & 180.3 & 103.66(42) & 3&184(4) & 167.5 & 65.11(29) & 3&129(5) & 177.5 & 63.63(12) & 4&5348(18)\\
			$\nu_2$ & 5&3512820(26) & 76.7 &  8.50(12) & 6&177(14) & 34.6 & 18.24(41) & 6&162(22) & 41.6 & 23.94(43) & 6&098(18) & 38.2 & 14.85(30) & 6&096(20) & 39.3 & 14.08(12) & 3&609(9)\\
			2$\nu_1$ & 10&3215360   & 39.1 &  4.34(12) & 2&754(27) & 13.6 & 7.18(42) & 2&67(6) & 16.8 & 9.64(44) & 2&56(5) & 16.8 & 6.53(31) & 2&51(5) & 19.0 & 6.80(12) & 5&309(18)\\
			$\nu_3$ & 8&376999(8)   & 25.5 &  2.83(12) & 2&20(4) & --- & --- & \multicolumn{2}{c}{---} & --- & --- & \multicolumn{2}{c}{---} & --- & --- & \multicolumn{2}{c}{---} & --- & ---&\multicolumn{2}{c}{---}\\
			$\nu_4$ & 5&284892(9)   & 23.4 &  2.60(12) & 4&63(5) & 8.6 & 4.55(45) & 1&62(10) & 11.1 & 6.40(48) & 1&43(8) & 10.0 & 3.88(34) & 1&61(9) & 12.4 & 4.44(12) & 3&524(28)\\
			$\nu_1+\nu_2$ & 10&5120499 & 17.5 &  1.94(12) & 2&72(6) & 7.7 & 4.07(40) & 5&64(10) & 9.5 & 5.49(43) & 5&34(8) & 8.9 & 3.46(30) & 5&33(9) & 9.5 & 3.40(12) & 4&36(4)\\
			$\nu_5$ & 8&596409(13)  & 16.2 &  1.80(12) & 4&62(7) & 10.8 & 5.71(44) & 2&71(8) & 13.9 & 7.98(47) & 2&45(6) & 12.6 & 4.89(33) & 2&50(7) & 6.7 & 2.39(12) & 5&16(5)\\
			$\nu_6$ & 11&603991(13) & 16.1 &  1.78(12) & 0&86(7) & 10.2 & 5.36(44) & 0&52(8) & 10.7 & 6.13(47) & 0&54(8) & 9.7 & 3.76(33) & 0&50(9) & --- & ---&\multicolumn{2}{c}{---}\\
			$\nu_7$ & 4&739179(13)  & 15.6 &  1.73(12) & 0&02(7) & 8.7 & 4.57(40) & 1&99(9) & 9.3 & 5.37(43) & 1&87(8) & 9.3 & 3.60(30) & 2&05(8) & 7.4 & 2.67(12) & 2&04(5)\\
			$\nu_8$ & 8&799128(21)  & 9.7 &  1.07(12) & 6&18(11) & 9.3 & 4.90(40) & 2&09(8) & 12.2 & 7.04(42) & 1&93(6) & 11.5 & 4.45(30) & 1&99(7) & --- & ---&\multicolumn{2}{c}{---}\\
			$\nu_9$ & 8&292173(25)  & 8.1 &  0.89(12) & 3&93(13) & 7.5 & 3.94(47) & 5&81(12) & 8.1 & 4.64(50) & 5&73(11) & 6.6 & 2.57(35) & 5&76(14) & --- & ---&\multicolumn{2}{c}{---}\\
			$\nu_{10}$ & 4&828367(27) & 7.5 &  0.83(12) & 1&83(14) & --- & --- &\multicolumn{2}{c}{---} & --- & --- &\multicolumn{2}{c}{---}& --- & --- &\multicolumn{2}{c}{---}& --- & ---&\multicolumn{2}{c}{---}\\
			$\nu_{11}$ & 8&35574(3) & 7.0 &  0.78(12) & 2&52(15) & --- & --- &\multicolumn{2}{c}{---}& --- & --- &\multicolumn{2}{c}{---}& --- & --- &\multicolumn{2}{c}{---}& --- & ---&\multicolumn{2}{c}{---}\\
			$\nu_1+\nu_3$ & 13&537767 & 5.5 &  0.61(12) & 4&14(19) &--- & --- &\multicolumn{2}{c}{---}& --- & --- &\multicolumn{2}{c}{---}& --- & --- &\multicolumn{2}{c}{---}& --- & ---&\multicolumn{2}{c}{---}\\
			$\nu_{12}$ & 4&88426(4) & 5.5 &  0.61(12) & 5&70(19) & --- & --- &\multicolumn{2}{c}{---}& --- & --- &\multicolumn{2}{c}{---}& --- & --- &\multicolumn{2}{c}{---}& --- & ---&\multicolumn{2}{c}{---}\\
			$\nu_{13}$ & 8&63759(5) & 4.9 &  0.55(12) & 5&66(21) & --- & --- &\multicolumn{2}{c}{---}& --- & --- &\multicolumn{2}{c}{---}& --- & --- &\multicolumn{2}{c}{---}& 5.0 & 1.78(12) & 0&06(7)\\
			$\nu_{14}$ & 11&62159(5) & 4.4 &  0.49(12) & 1&95(24) & --- & --- &\multicolumn{2}{c}{---}& --- & --- &\multicolumn{2}{c}{---}& --- & --- &\multicolumn{2}{c}{---}& --- & ---&\multicolumn{2}{c}{---}\\
			3$\nu_1$ & 15&4823040  & 4.2 &  0.46(12) & 5&28(25) & --- & --- &\multicolumn{2}{c}{---}& --- & --- &\multicolumn{2}{c}{---}& --- & --- &\multicolumn{2}{c}{---}& --- & ---&\multicolumn{2}{c}{---}\\
			$\nu_{15}$ & 8&312912(7) & --- & --- & \multicolumn{2}{c}{---} &  --- & --- &\multicolumn{2}{c}{---}& --- & --- &\multicolumn{2}{c}{---} & --- & --- &\multicolumn{2}{c}{---}& 10.2 & 3.66(13) & 5&20(4)\\
			\hline
			SDR [mmag] & \multicolumn{2}{c}{} && 12.35 &\multicolumn{2}{c}{}&& 6.50 &\multicolumn{2}{c}{}&& 6.68 &\multicolumn{2}{c}{}&& 4.65 &\multicolumn{2}{c}{}&& 6.39\\
			DT [mmag] & \multicolumn{2}{c}{} && 0.44 &\multicolumn{2}{c}{}&& 2.11 &\multicolumn{2}{c}{}&& 2.30 &\multicolumn{2}{c}{}&& 1.56 &\multicolumn{2}{c}{}&& 1.43\\
			$N_{\rm obs}$ & \multicolumn{2}{c}{} && 22795 &\multicolumn{2}{c}{}&& 558 &\multicolumn{2}{c}{}&& 564 &\multicolumn{2}{c}{}&& 559 &\multicolumn{2}{c}{}&& 6504 \\
			\hline
			\label{tab:freqs}
		\end{tabular}
	\end{table}
\end{landscape}

\subsection{Re-analysis of Templeton et al.'s data}\label{sect:templ}
Having derived the frequencies from the SMEI data, we re-analyzed the $y$-filter data published by \cite{Templeton1997} and kindly provided by Dr.~Templeton. The data were obtained in the years 1983\,--\,1988 in three sites; we shall refer to them as to the `MTy data'. Due to the large gaps in the data, they suffer from strong aliases as shown, for example, in their fig.\,3. We recovered all eight frequencies listed in their Table 1. However, two of the frequencies reported by them are slightly deviant from our solution.
These were 54.82807\,$\mu$Hz (4.737145\,d$^{-1}$), which differs from $\nu_7$ by about 0.0020~d$^{-1}$ and 99.44722\,$\mu$Hz (8.592240\,d$^{-1}$), which is off by about 0.0042\,d$^{-1}$ from $\nu_5$. This is the result of identifying an alias as the true frequency.
This possibility was pointed out by \cite{Templeton1997}, who wrote that frequencies of the lower-amplitude peaks they detected could be in error by 0.016 to 0.058\,$\mu$Hz. This corresponds to 0.0014 - 0.0050\,d$^{-1}$, in full accordance with the above differences. In addition to the eight confirmed terms, we also detected $\nu_{13}$, one of the smallest-amplitude frequency found in the SMEI data. Amplitudes of all 9 terms detected in MTy data are given in Table \ref{tab:freqs}.

\begin{figure}
\includegraphics[width=\columnwidth]{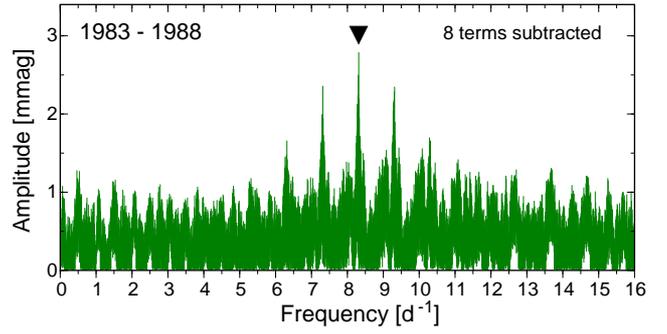}
\caption{Fourier frequency spectrum from the $y$-filter data of \protect\cite{Templeton1997} after subtraction of all terms detected in these data except for $\nu_{15}$. The highest peak corresponding to $\nu_{15}$ is marked by the inverted triangle.}
\label{fig:templ-fs}
\end{figure}

A comparison of the frequency values (and their amplitudes) detected in the data of \cite{Templeton1997} with those found in the SMEI and APT data provides additional evidence of the changes of amplitudes of the independent frequencies. For example, the amplitude of $\nu_5$ is much smaller in the MTy data than in the APT data. Despite similar detection thresholds, $\nu_6$, $\nu_8$, and $\nu_9$ present in the APT data, were not detected in the MTy data. On the other hand, the frequency $\nu_{15}$ was detected only in the MTy data. Its reality is demonstrated in Fig.\,\ref{fig:templ-fs}, in which we show the frequency spectrum of MTy data freed from all frequencies detected in these data but $\nu_{\rm 15}$. The strongest alias of this frequency is about 10\% weaker, which leaves little ambiguity as to the correctness of the frequency $\nu_{15}$.

 The detection of  $\nu_{15}$ only in the MTy data and a lack of $\nu_{13}$ in  the APT data is again a manifestation of the amplitude modulation that occurs quite often in $\delta$ Scuti variables \citep[e.g.,][]{BregerPam2006, Bowman2016}.

\section{Mode identification}

Here, we use the method of mode identification which is based on simultaneous determination of the mode degree $\ell$,
the intrinsic mode amplitude $\varepsilon$ and the non-adiabatic parameter $f$ for a given frequency \citep{JDD2003, JDD2005}.
This method consists in least-squares fitting of the calculated values of the complex photometric amplitudes to the observed ones.
We regard the degree $\ell$  and associated complex values of $f$ and $\varepsilon$ as most probable
if they minimise the differences between the computed and observed complex amplitudes. Then, the empirical values of $f$
can be directly compared with predictions from linear computations of stellar pulsations.
\begin{figure*}
	\includegraphics[width=175mm,clip]{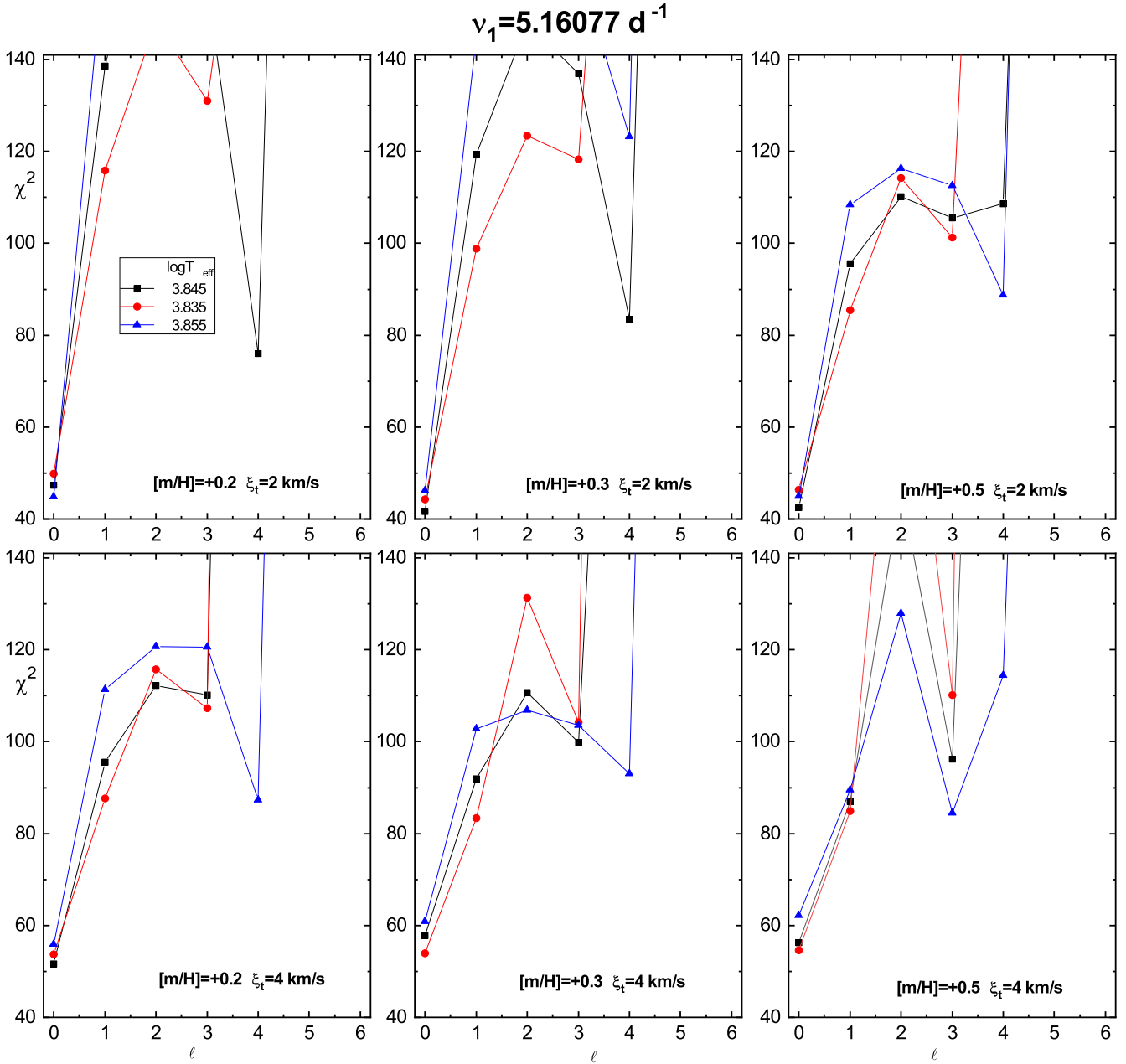}
	\caption{The discriminant $\chi^2$ as a function of $\ell$ for the dominant frequency $\nu_1$ for various combinations
of  [m/H] and $\xi_t$.}
	\label{fig6}
\end{figure*}

Below, we repeat basic formulae to remind the meaning of $\varepsilon$ and $f$.
The standard formula for the relative local radial displacement of the surface element caused by a pulsational mode
with the angular frequency $\omega$ is
$$\delta r(R,\theta,\varphi)= R {\rm Re}\{ \varepsilon Y_\ell^m(\theta,\varphi)
{\rm e}^{-{\rm i}\omega t}\}, \eqno(1)$$
where $Y_{\ell}^m$ denotes a spherical harmonic with degree $\ell$ and azimuthal order $m$ and $G,M,R$ have their usual meanings.
The corresponding changes of the bolometric flux, ${\cal F}_{\rm bol}$ are given by
$$\frac{ \delta {\cal F}_{\rm bol} } { {\cal F}_{\rm bol} }= {\rm
Re}\{ \varepsilon f Y_\ell^m (\theta,\varphi) {\rm e} ^{-{\rm i} \omega t} \}.\eqno(2)$$
Eq.\,(2) defines the complex parameter $f$ as the ratio of the relative flux variation
to the relative radial displacement of the surface.

Both $\varepsilon$ and $f$ may be regarded constant in the atmosphere, so we can use the static plane-parallel approximation.
Thus, in the framework of linear theory of stellar pulsations, assuming the zero-rotation approach,
the complex amplitude of the relative flux variation in a passband $\lambda$ for a given pulsational mode can be
written as follows \citep[e.g.][]{JDD2003, JDD2005}:
$${\cal A}^{\lambda}(i) = {\cal D}_{\ell}^{\lambda} ({\tilde\varepsilon}
f) +{\cal E}_{\ell}^{\lambda} {\tilde\varepsilon}, \eqno(3)$$
where
$${\tilde\varepsilon}\equiv \varepsilon Y^m_{\ell}(i,0),\eqno(4a)$$
$${\cal D}_{\ell}^{\lambda} = b_{\ell}^{\lambda} \frac14
\frac{\partial \log ( {\cal F}_\lambda |b_{\ell}^{\lambda}| ) }
{\partial\log T_{\rm{eff}}}, \eqno(4b)$$
$${\cal E}_{\ell}^{\lambda}= b_{\ell}^{\lambda} \left[ (2+\ell
)(1-\ell ) -\left( \frac{\omega^2 R^3}{G M}
 + 2 \right) \frac{\partial \log ( {\cal F}_\lambda
|b_{\ell}^{\lambda}| ) }{\partial\log g} \right],\eqno(4c)$$
and
$$b_{\ell}^{\lambda}=\int_0^1 h_\lambda(\mu) \mu P_{\ell}(\mu) d\mu.\eqno(4d)$$
The term ${\cal D}_{\ell}^\lambda$ describes the temperature effects and ${\cal E}_{\ell}$ combines
the geometrical and pressure effects.
The partial derivatives of ${\cal F}_\lambda(T_{\rm eff},\log g)$ and $b_{\ell}^{\lambda}(T_{\rm eff},\log g)$
may be calculated numerically from tabular data. Their values depend also on the metallicity
parameter [m/H] and microturbulent velocity $\xi_t$ in the atmosphere.
In this paper we use Vienna model atmospheres \citep{Heiter2002} that include turbulent convection treatment from \citet{Canuto1996}.
Although the atmosphere of $\delta$ Scuti shows some chemical peculiarity, as mentioned in Sect.\,2,
the abundances of elements with the atomic number $Z < 27$ are quite close to the solar values \citep{Yushchenko2005}.
Moreover, here we do not analyse variations in single spectral lines but consider the light variations in the photometric passbands
which have the width of about 200-300\AA. To give some quantitative measurement of the chemical peculiarity in the photometric bands,
we calculated the peculiarity index $\Delta p$ based on the Str\"omgren photometry as given by \citet{Masana1998}.
Using the photometric indices from \citet{Paunzen2015} we got $\Delta p=1.11$ which is lower than the threshold for normal stars of 1.50.
Thus, it is reasonable to use the standard Vienna atmospheres for computations of the photometric amplitudes and phases
of the $\delta$ Scuti pulsations.

For the limb darkening law, $h_\lambda(\mu)$, we computed coefficients assuming the nonlinear, four-parametric formula of \citet{Claret2000}. The symbol $i$ in Eq.\,(4a) means the inclination angle.
To convert the flux amplitudes to magnitudes, the right hand side of Eq.\,(3) must be multiplied by the factor $(-1.086)$.
\begin{figure*}
	\includegraphics[width=175mm,clip]{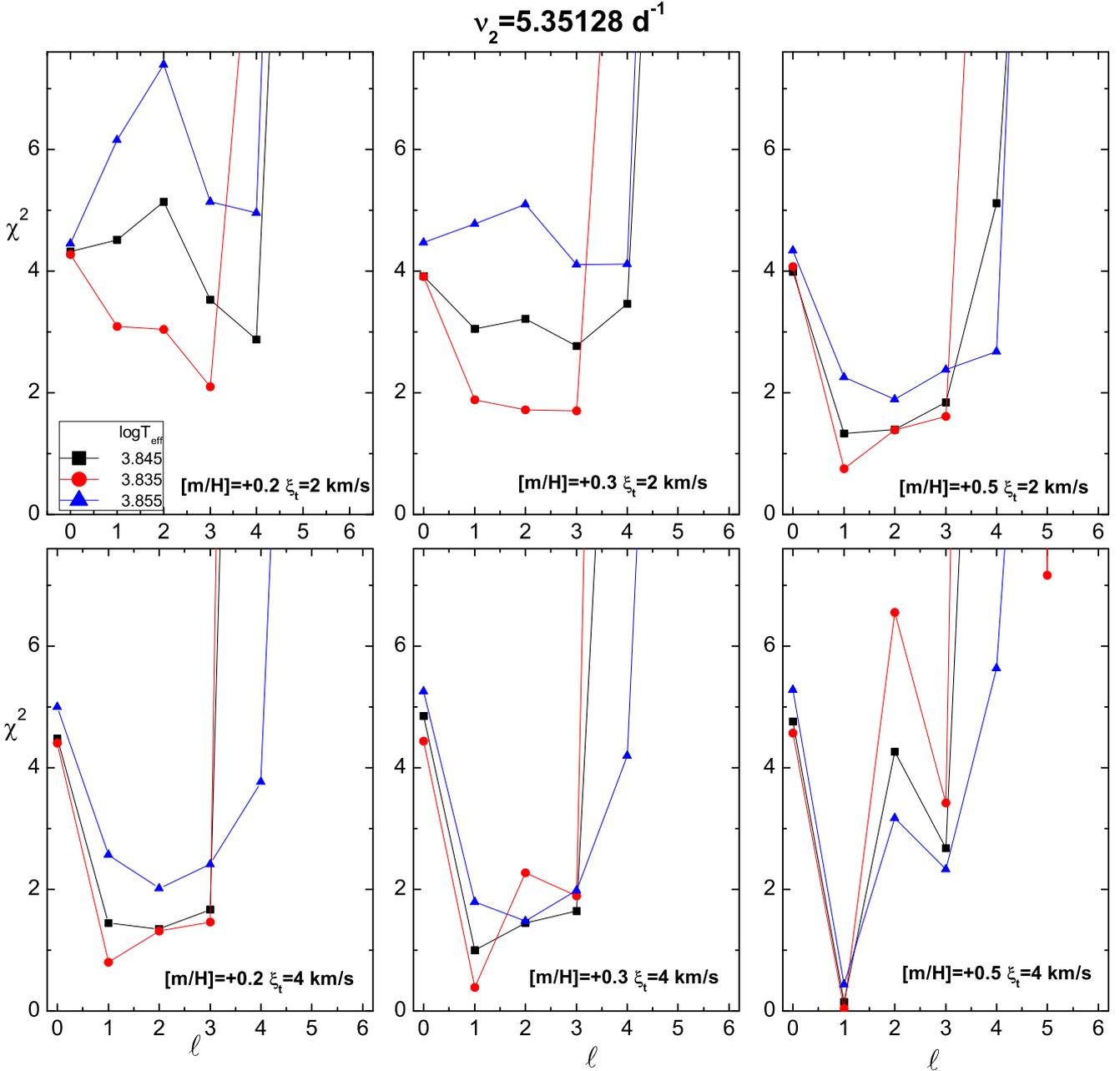}
	\caption{The same as in Fig.\,6 but for the frequency $\nu_2$.}
	\label{fig7}
\end{figure*}

The goodness of the fit is obtained from the formula
$$\chi^2=\frac1{2N-N_p} \sum_{i=1}^N  \frac{ |{\cal A}^{obs}_{\lambda_i} - {\cal A}^{cal}_{\lambda_i}|^2 }{ |\sigma_{\lambda_i}|^2},\eqno(5)$$
where $N$ is the number of passbands $\lambda_i$ and $N_p$ is the number of parameters to be determined.
The method yields two complex parameters, $\tilde\varepsilon$ and $f$, thus $M=4$.
The symbols ${\cal A}^{obs}$ and ${\cal A}^{cal}$ denote the complex observational and calculated amplitudes,
respectively. The observational errors $\sigma_{\lambda}$ are expressed as
$$|\sigma_\lambda|^2= \sigma^2 (A_{\lambda})  +  A_{\lambda}^2 \sigma^2(\varphi_\lambda), \eqno(6)$$
where $A_{\lambda}=|{\cal A}_\lambda|$ and $\varphi_{\lambda}=arg({\cal A}_\lambda)$,
are the values of the amplitude and the phase, respectively.
\begin{figure*}
	\includegraphics[width=175mm,clip]{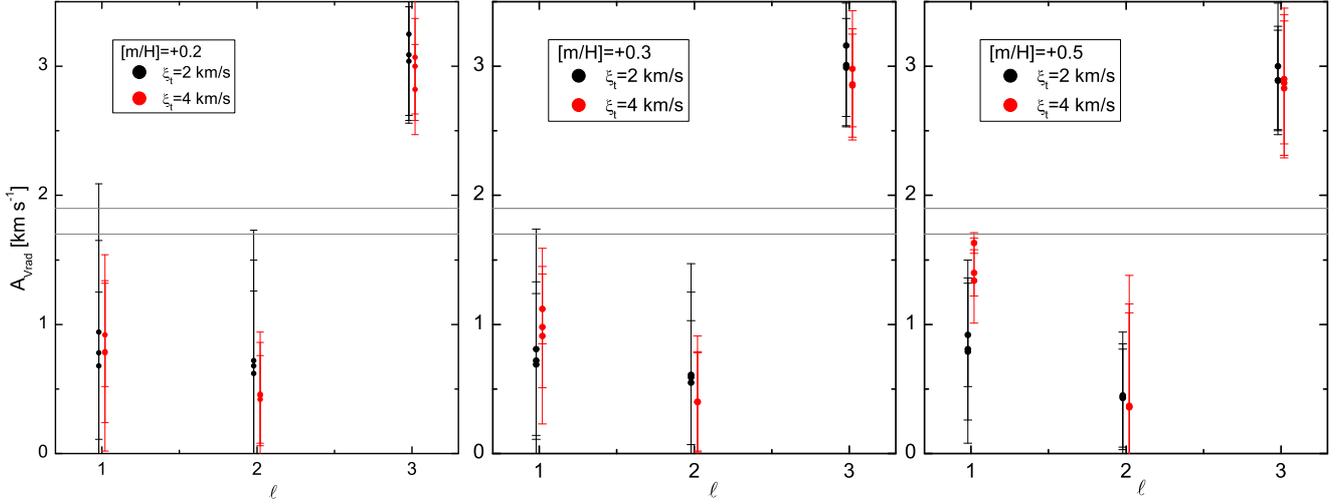}
	\caption{The estimated amplitude of the radial velocity for the second frequency $\nu_2$ as a function of the possible mode degrees $\ell$.
The panels from the left to the right show the results for $\mbox{[m/H]}=0.2, ~0.3$ and 0.5, for the two values of $\xi_t$ in each case.
The same three values of $T_{\rm eff}$ as in Figs.\, 6 and 7 were considered.
The horizontal lines mark the observed range as determined by \citet{Balona1981}.}
	\label{fig8}
\end{figure*}

As has been mentioned in Sect.\,2, the metallicity of $\delta$ Scuti is in the range of about $\mbox{[m/H]}\in(0.2,~0.4)$
and the microturbulent velocity in its atmosphere is about $\xi_t=2.8~\kms$.
In addition, the prototype has been classified as a chemically peculiar star because of an excess of some elements, e.g.,
iron, nickel and manganese. Therefore, we consider three values of the atmospheric metallicity which are in the tabular data
$\mbox{[m/H]}=0.2,~0.3$ and 0.5 and two values of the microturbulent velocity $\xi_t=2$ and 4$~\kms$, which gives six combinations of ([m/H],\,$\xi_t$).
In Fig.\,6, we plotted the values of $\chi^2$ as a function of the mode degree $\ell$ for the dominant frequency $\nu_1=5.16077$\,d$^{-1}$
considering the six pairs of ([m/H],\,$\xi_t$) and the three values of the effective temperature $\log T_{\rm eff}=3.835, 3.845$ and 3.855
that cover the observed range.
The value of luminosity in the adopted range, $\log L/L_{\sun}= 1.578(17)$, has only a minor effect on $\chi^2(\ell)$.
The plotted values of the discriminant are for $\log L/L_{\sun}= 1.578$.
As one can see, for all values of $\log T_{\rm eff}$ and all pairs of ([m/H],\,$\xi_t$), a radial mode is strongly preferred.
However, the run of $\chi^2(\ell)$ is very sensitive to the effective temperature as well as to the atmospheric parameters ([m/H],\,$\xi_t$).
Even for fixed values of ([m/H],\,$\xi_t$) we have a very different dependence $\chi^2(\ell)$
for different values of $\log T_{\rm eff}$. Only in the case of the metallicity $\mbox{[m/H]}=0.5$ and microturbulent velocity $\xi_t=4~\kms$
the function $\chi^2(\ell)$ has a similar course for the three values of $\log T_{\rm eff}$.

Because in case of radial modes  $Y^m_{\ell}(i,0)=1$, our method gives an exact value of the intrinsic mode amplitude $\varepsilon$
which for ([m/H],\,$\xi_t$) =(0.5,\,4$~\kms$) is $\varepsilon=0.006(3)$. This means that the pulsations of the dominant mode change
the radius by 0.3-0.9\,\% at the photosphere level.

The amplitude of the radial velocity for a given pulsational mode can be estimated from the formula
$${\cal A}(V_{\rm rad})= {\rm i}\omega R \left( u_{\ell}^{\lambda}
+ \frac{GM}{R^3\omega^2} v_{\ell}^{\lambda} \right) \tilde\varepsilon,\eqno(7)$$
taking the empirical value of $\tilde\varepsilon= \varepsilon Y^m_{\ell}(i,0)$ obtained from our method.
The factors $u_{\ell}^{\lambda}$ and $v_{\ell}^{\lambda}$ are other integrals of the limb darkening law $h_\lambda(\mu)$
$$u_{\ell}^{\lambda} = \int_0^1 h_\lambda(\mu) \mu^2 P_{\ell}(\mu) d\mu,\eqno(7a)$$
$$v_{\ell}^{\lambda} = \ell \int_0^1 h_\lambda(\mu) \mu \left(P_{\ell-1}(\mu) - \mu P_{\ell}(\mu) \right)d\mu, \eqno(7b)$$
and are computed for the Str\"omgren passband $y$.
\begin{figure*}
	\includegraphics[width=175mm,clip]{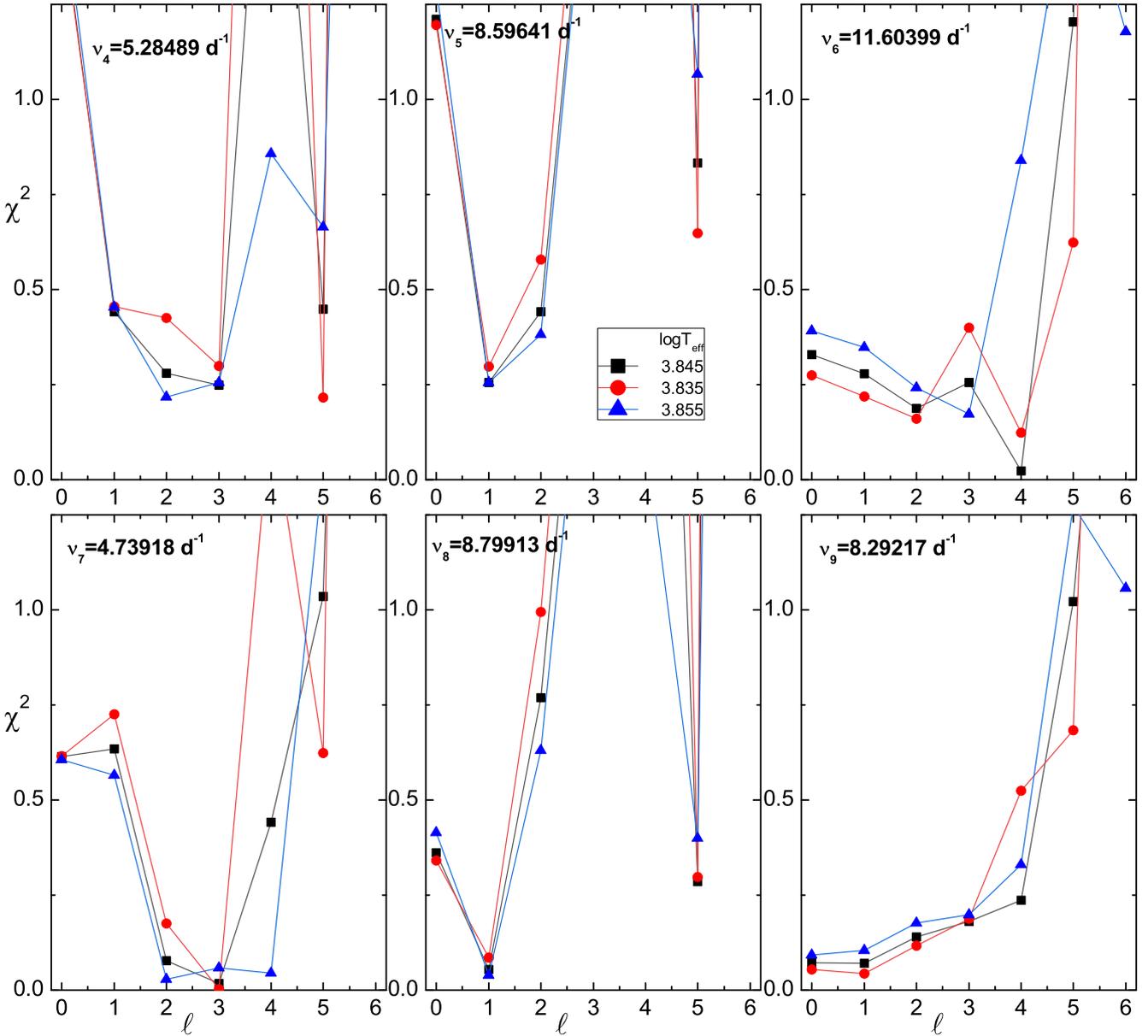}
	\caption{The value of $\chi^2(\ell)$ for the remaining six frequencies of $\delta$ Scuti for the atmospheric metallicity [m/H]=+0.5
       and the microturbulent velocity $\xi_t=4~\kms$. The same values of the effective temperature as in Figs.\,6 and 7 were considered.}
	\label{fig9}
\end{figure*}

The latest determination of the radial velocity amplitudes for the pulsation frequencies of $\delta$ Scuti is due to \citet{Balona1981}.
They fitted all the published radial velocities using the two main frequencies and the first harmonic of the dominant one
and obtained the amplitudes: $A_{\rm Vrad}=4.1(1)~\kms$ for $\nu_1$ and $A_{\rm Vrad}=1.8(1)~\kms$ for $\nu_2$.
Our estimate for $\nu_1$ calculated from Eq.\,(7) is about $A_{\rm Vrad}=4.8(2.6)~\kms$ and it agrees within the errors with the value of \citet{Balona1981}.
Of course, the amplitudes could change within about 40 years between the two data sets, but comparing the amplitude in the $V$ filter of \citet{Balona1981}
with our $y$ amplitude we conclude that they agree within 2.5$\sigma$.

In Fig.\,7, we presented the results of the $\ell$ identification for the second frequency $\nu_2=5.35128$\,d$^{-1}$.
The light amplitude of $\nu_2$ is about 4.5 times smaller than the amplitude of the dominant frequency.
In the case of $\nu_2$, one can see even a greater dependence of the function $\chi^2(\ell)$ on the parameters ([m/H],\,$\xi_t$)
and on $T_{\rm eff}$ and, most importantly, the uniqueness of the identification of $\ell$ depends on them.
For most combinations of ([m/H], $\xi_t$) the degrees $\ell=1,2,3$ are equally possible.
As in the case of $\nu_1$, a similar run of $\chi^2(\ell)$ for each $T_{\rm eff}$ is obtained for ([m/H],\,$\xi_t$)=(0.5, 4$~\kms$).
For these parameters, we also obtain an unambiguous identification of the degree $\ell=1$ .

However, in order to definitely decide on the identification of $\ell$ for $\nu_2$, we calculated the amplitude of the radial velocity according
to Eq.\,(7). In Fig.\,8, we show these empirical values of $A_{\rm Vrad}=|{\cal A}(V_{\rm rad})|$ as a function of $\ell$,
computed assuming the intrinsic amplitudes $\tilde\varepsilon$ as determined simultaneously with $\ell$ and the parameter $f$.
All combinations of ([m/H,\,$\xi_t$]) depicted in Fig.\,7 were considered and the same three values of $T_{\rm eff}$  were assumed.
The horizontal lines mark the allowed observed range as determined by \citet{Balona1981}.
As one can see, only for a dipole mode our estimated values of $A_{\rm Vrad}$ reach the observed range
for all values of the metallicity [m/H] and microturbulent velocity $\xi_t$.
Thus, even though the radial velocity amplitude can slightly change, we conclude that an $\ell=1$ mode is the most probable
identification for frequency $\nu_2$.

The discriminant $\chi^2(\ell)$ for the remaining six frequencies is plotted in Fig.\,9, for the atmospheric parameters [m/H]=0.5,\,$\xi_t=4~\kms$
which gave the most homogeneous dependence of $\chi^2(\ell)$ on the effective temperature for $\nu_1$ and $\nu_2$. We remind that the frequency $\nu_3=8.376999$\,d$^{-1}$
was not detected in the APT data, therefore the photometric identification of the degree $\ell$ is not possible for it.
As one can see in Fig.\,9, the frequency $\nu_4=5.28489$\,d$^{-1}$ can be  $\ell=1,2,3$ or 5, however $\ell=5$ is much less probable because its visibility is much lower.
The frequency $\nu_5=8.59641$\,d$^{-1}$ is most probably a mode with $\ell=1$ or 2. In the case of $\nu_6=11.60399$\,d$^{-1}$ we can exclude modes with $\ell>4$
as well as a radial mode because there are no radial overtones in pulsational models with close frequencies.
The most likely mode degrees for $\nu_7=4.73918$\,d$^{-1}$ are $\ell=2,3,4$, but $\ell=4$ is indicated only by the hottest effective temperature.
The frequency  $\nu_8=8.79913$\,d$^{-1}$ is most probably the mode $\ell=1$,
because $\ell=5$ has much lower visibility and $\ell=0$ is excluded because no theoretical radial overtones have frequencies close to it.
In the case of $\nu_9=8.29217$\,d$^{-1}$ we can only say that this is a mode with $\ell\le 4$ and exclude $\ell=0$ because of the frequency value.

\section{Models fitting the dominant frequency as the radial mode}

Pulsational models for $\delta$ Scuti were computed with the linear nonadiabatic code of \citet{Dziembowski1977a}.
The code takes into account the effects of rotation up to the second order.
The convective flux is assumed to be constant during the pulsational cycle; this is called the convective flux freezing approximation.
This approximation is adequate if convection is not very efficient, i.e., if it does not dominate the energy transport.
In the case of the efficient convection, mode properties and driving can be affected in $\delta$ Sct star models.
However, as we showed on the example of another $\delta$ Sct star FG Vir with similar parameters to $\delta$ Scuti,
the results from different treatments of convection are similar for $\alpha_{\rm MLT}$ less than about 1.0 \citep{JDD2005}.

Three opacity tables were used: OPAL \citep{Iglesias1996}, OP \citep{Seaton2005} and OPLIB \citep{Colgan2015, Colgan2016}.
In the first step, we searched for models that reproduce the dominant frequency $\nu_1=5.16077$\,d$^{-1}$ as a radial mode.
Fig.\,10 shows evolutionary tracks with such models marked on them. The dots represent the radial fundamental mode ($n=1$)
and the triangles - the radial first overtone ($n=2$). The models were computed with OPAL opacities but
the results obtained with OP and OPLIB opacity data are qualitatively the same and quantitatively not very different.
The four panels show the results computed for various combinations of  $X_0$, $Z$ and $\alpha_{\rm ov}$, adopting $\alpha_{\rm MLT}=0.5$.
For the considered masses and effective temperatures, the effect of $\alpha_{\rm MLT}$ is minor
and, for example, the models computed with $\alpha_{\rm MLT}=1.8$ that reproduce $\nu_1$ as the fundamental mode
have parameters that differ in value to the fourth decimal place.
As one can see, in all cases the first overtone is far outside the error box of $\delta$ Sct. Therefore, we unequivocally
conclude that the dominant frequency is the radial fundamental mode and thus we confirm the result of \citet{Templeton1997}.
\begin{figure*}
\begin{multicols}{2}
\includegraphics[width=88mm,clip]{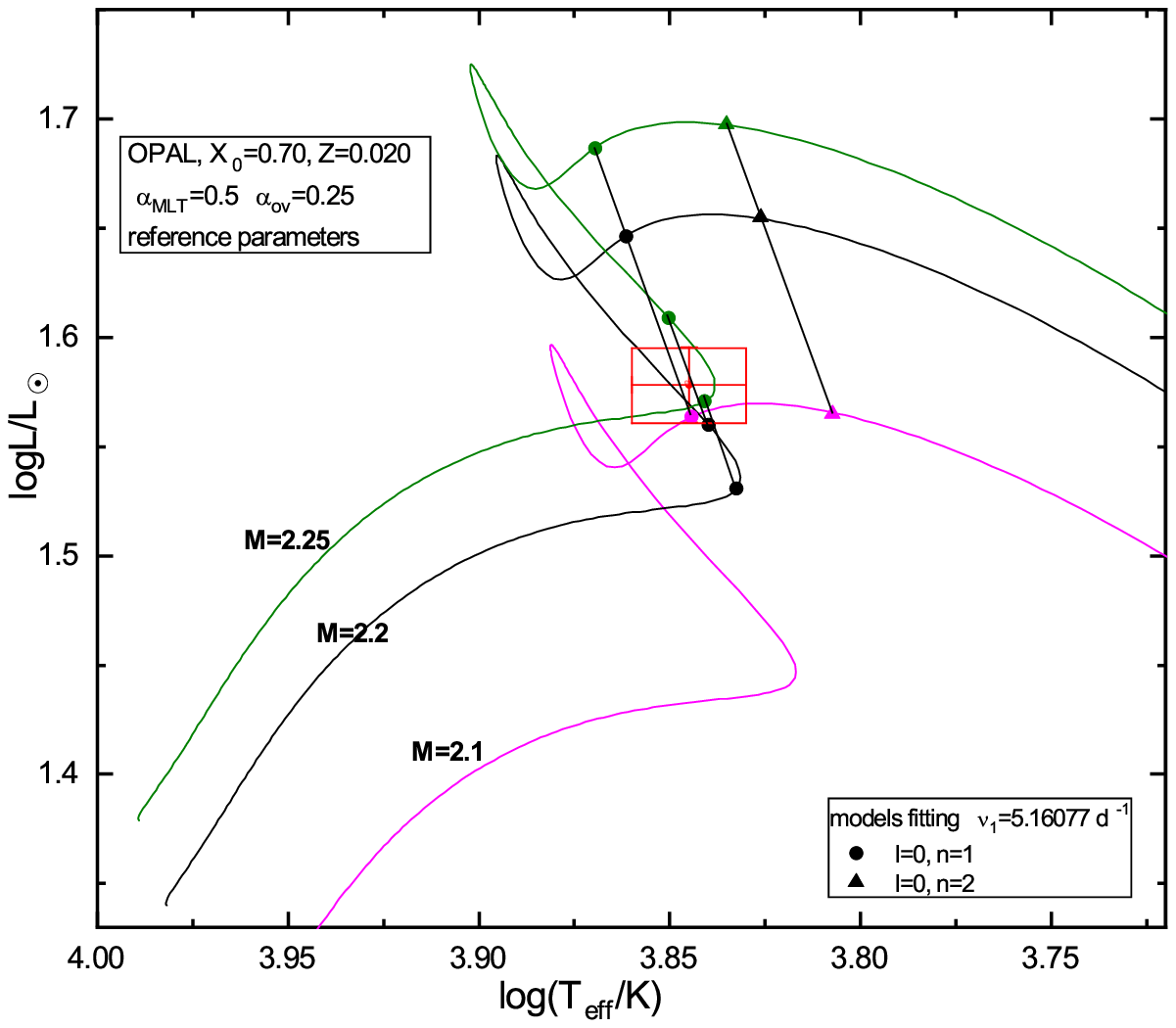}\par
\includegraphics[width=88mm,clip]{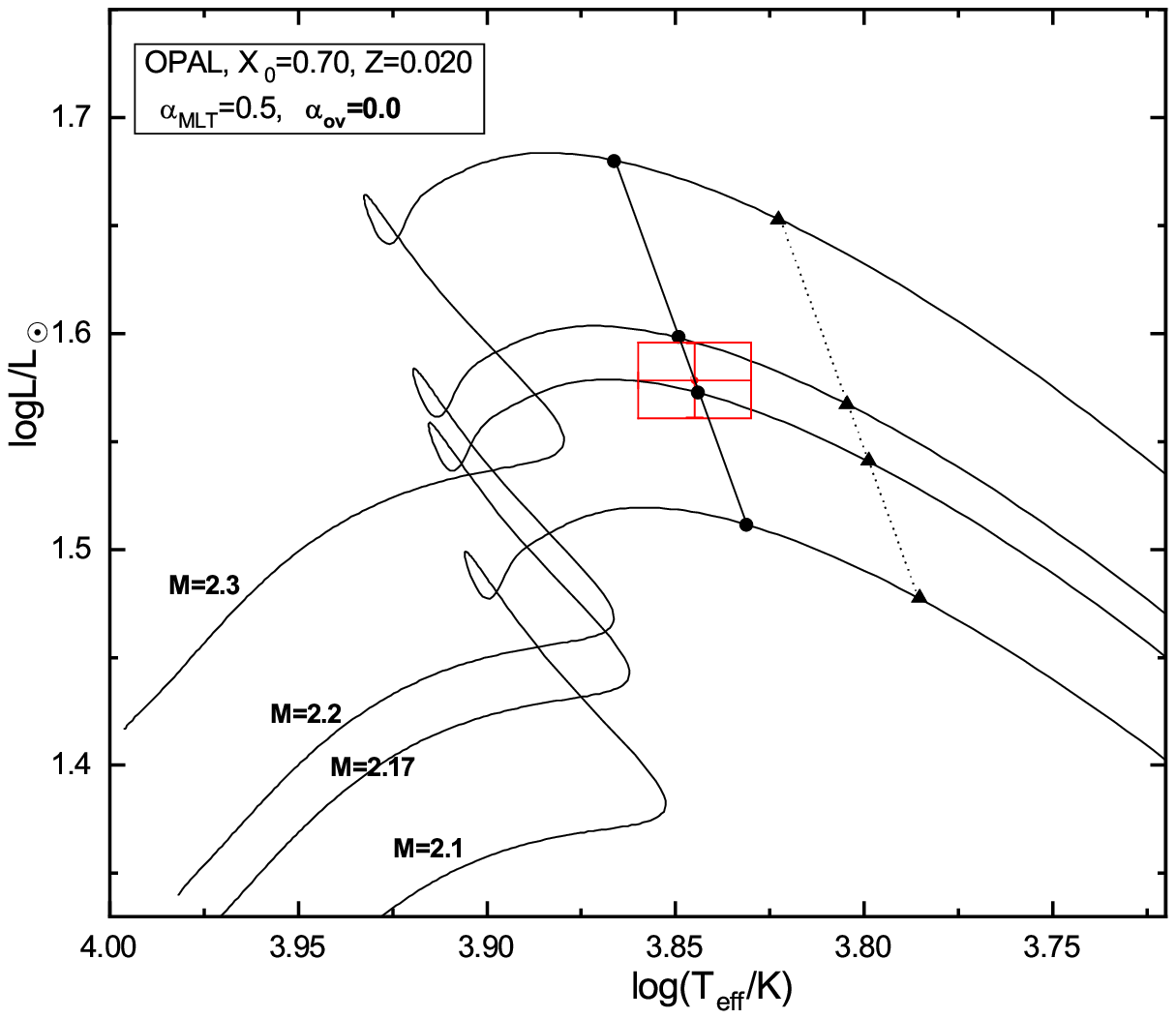}\par
\end{multicols}
\begin{multicols}{2}
\includegraphics[width=88mm,clip]{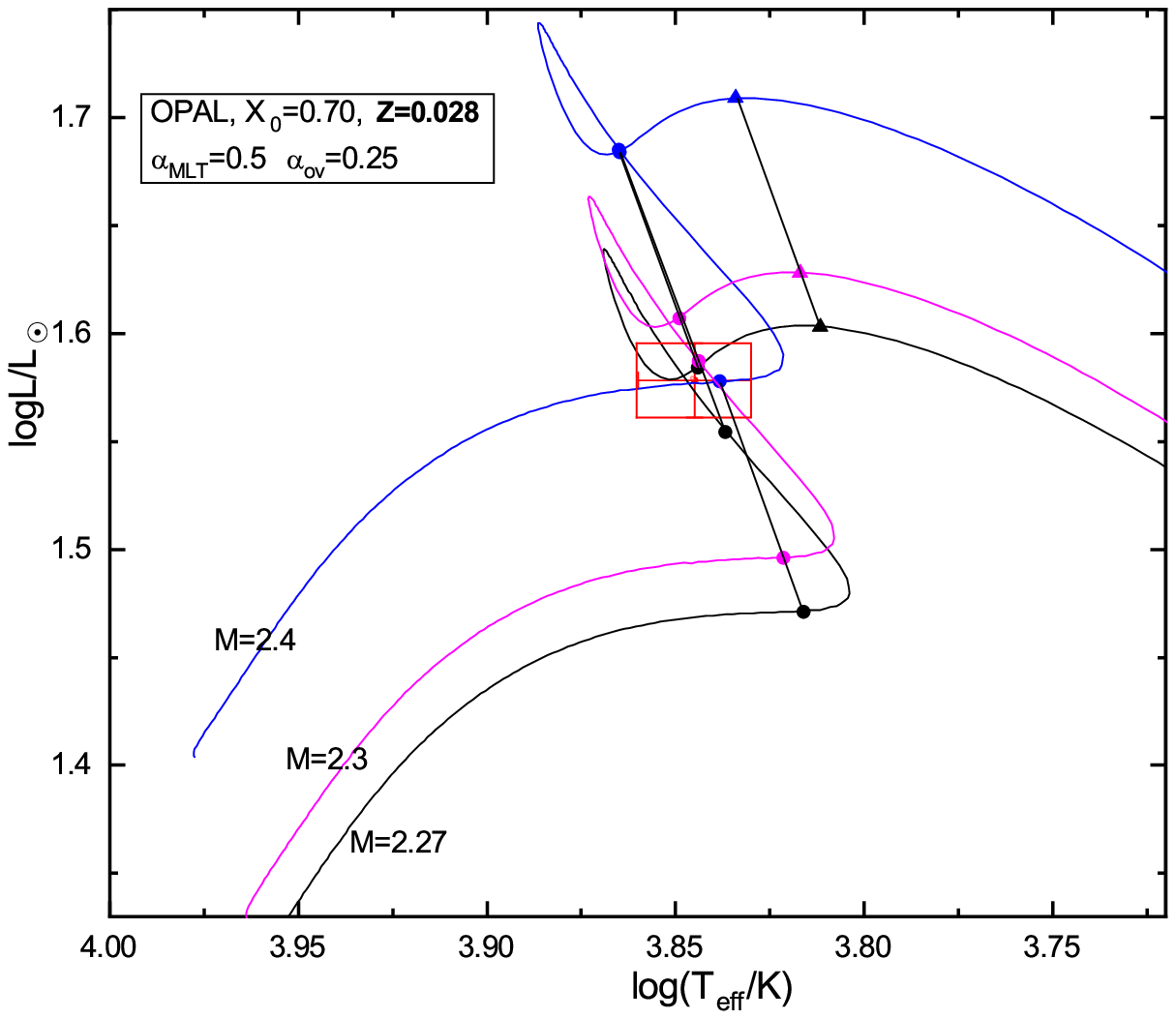}\par
\includegraphics[width=88mm,clip]{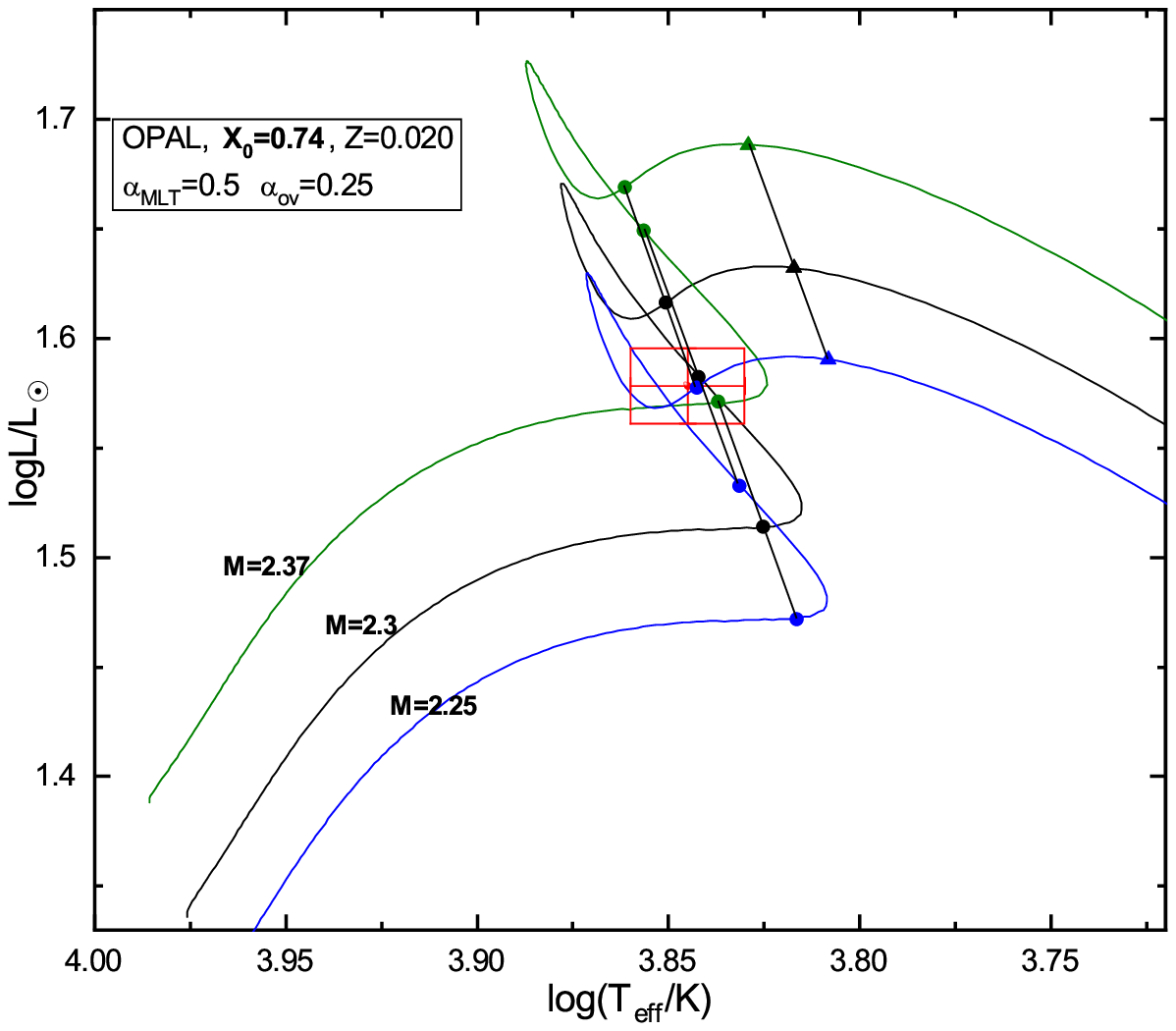}\par
\end{multicols}
\caption{The OPAL evolutionary tracks with models marked as dots or triangles that reproduce the dominant frequency of $\delta$ Scuti $\nu_1=5.16077$\,d$^{-1}$ as a radial mode. Dots correspond to the fundamental mode and triangles to the first overtone mode.
The observed parameters of $\delta$ Scuti are depicted together with their error box.
Each of the four panels show results for various combinations of the initial hydrogen amount $X_0$, metallicity $Z$
and overshooting from the convective core $\alpha_{\rm ov}$. A mixing length parameter
in the envelope of $\alpha_{\rm MLT}=0.5$ was adopted in each case. The top-right panel is for the no-overshooting case.}
\label{fig10}
\end{figure*}

Now, let us return to the question of the evolutionary stage of $\delta$ Scuti. In case of models computed without overshooting
from the convective core (the top-right panel of Fig.\,10), only the post-main sequence phase is allowed within the observational error box.
It is worth to add that in case of models with the metallcity lower than $Z=0.019$ at $\alpha_{\rm ov}=0.25$, the dominant mode reaches the observed value of the frequency also only in post-main sequence stage.
In case of models computed with overshooting from the convective core of $\alpha_{\rm ov}=0.25$ and $Z>0.019$ (the other three panels of Fig.\,10), the frequency of the radial fundamental mode can be reproduced in the three stages of evolution: MS, OC and post-MS.
Moreover, the models in these three stages can have very similar stellar parameters ($\log T_{\rm eff},~\log L/L_{\sun}$).
In Table\,2, we give the parameters of such models computed for various combinations of the parameters
$X,~Z,~\alpha_{\rm ov}$ for the two values of the mixing length parameter in the envelope $\alpha_{\rm MLT}=0.5$ and $\alpha_{\rm MLT}=1.8$.
In addition to these parameters, Table\,2 also includes:
mass, the phase of evolution, effective temperature,
luminosity, radius, the real and imaginary part of the non-adiabatic parameter $f$ of the radial fundamental mode
and the normalized work integral $\eta$. The positive value of $\eta$ means that the pulsational mode is unstable (i.e. excited).
%
\begin{table*}
\centering
\caption{The parameters of the OPAL models that reproduce $\nu_1=5.16077$\,d$^{-1}$ as the radial fundamental mode ($\ell=0,~n=1$).
The following parameters are given in the subsequent columns: the initial hydrogen abundance $X_0$, metallicity $Z$,
the overshooting from the convective core $\alpha_{\rm ov}$, mass, phase of evolution, age, effective temperature, luminosity, radius,
the real and imaginary part of the non-adiabatic parameter $f$ of the radial fundamental mode and the normalized instability parameter
$\eta$. Most of the results are for $\alpha_{\rm MLT}=0.5$ and, for a comparison, at the very bottom of the table we give the results for
$\alpha_{\rm MLT}=1.8$ at $X_0=0.70,~Z=0.020,~\alpha_{\rm ov}=0.25$. For $\alpha_{\rm ov}=0.0$ only the post-MS phase is possible.}
\begin{tabular}{ccccccccccccc}
\hline
\multicolumn{4}{c}{$\alpha_{\rm MLT}=0.5$} \\
\hline
$X_0$ & $Z$ & $\alpha_{\rm ov}$ & $M/M_{\sun}$ & phase & age [Gyr] & $\log T_{\rm eff}/{\rm K}$ & $\log L/L_{\sun}$ & $R/R_{\sun}$ & $f_R$ & $f_I$ & $\eta$ \\
\hline
 0.70 & 0.014 & 0.25 & 1.97 & post-MS & 1.0617 & 3.8475 & 1.560 & 4.07 &  0.757  &  $-12.217$ &  0.056\\
\hline
 0.70 & 0.017 & 0.25 & 2.04 & post-MS & 1.0574 & 3.8462 & 1.564 & 4.11 &  0.712  &  $-12.185$ &  0.054\\
\hline
 0.70 & 0.020 & 0.25 & 2.25 &    MS   & 0.8422 & 3.8408 & 1.571 & 4.25 &  2.324  &  $-10.656$ &  0.071\\
      &       &      & 2.20 &    OC   & 0.9218 & 3.8398 & 1.560 & 4.21 &  2.640  &  $-10.180$ &  0.075\\
      &       &      & 2.10 & post-MS & 1.0597 & 3.8443 & 1.564 & 4.14 &  1.070  &  $-11.840$ &  0.057\\
\hline
 0.70 & 0.020 & 0.00 & 2.17 & post-MS & 0.8218 & 3.8440 & 1.573 & 4.19 &  1.171  &  $-11.845$ &  0.057\\
\hline
 0.70 & 0.028 & 0.25 & 2.40 &   MS    & 0.8154 & 3.8383 & 1.578 & 4.33 &  2.253  & $-10.770$ &  0.067\\
      &       &      & 2.30 &   OC    & 0.9859 & 3.8439 & 1.587 & 4.27 &  0.178  &  $-12.657$ &  0.042\\
      &       &      & 2.27 & post-MS & 1.0252 & 3.8441 & 1.584 & 4.25 &  0.038  &  $-12.725$ &  0.041\\
\hline
 0.74 & 0.020 & 0.25 & 2.37 &  MS     & 0.8878 & 3.8370 & 1.571 & 4.32 &  2.781  &  $-9.988$ &  0.074\\
      &       &      & 2.30 &  OC     & 1.0178 & 3.8422 & 1.582 & 4.28 &  0.964  &  $-12.185$ &  0.052\\
      &       &      & 2.25 & post-MS & 1.0857 & 3.8426 & 1.578 & 4.24 &  0.778  &  $-12.277$ &  0.050\\
\hline
\multicolumn{4}{c}{$\alpha_{\rm MLT}=1.8$} \\
\hline
 0.70 & 0.020 & 0.25 & 2.25 &   MS    & 0.8422 & 3.8411 & 1.571 & 4.24 &  2.837  &   $-0.172$ &  0.104\\
      &       &      & 2.20 &   OC    & 0.9218 & 3.8403 & 1.560 & 4.20 &  2.705  &   $-0.141$ &  0.108\\
      &       &      & 2.10 & post-MS & 1.0597 & 3.8446 & 1.564 & 4.14 &  3.329  &   $-0.465$ &  0.096\\
\hline
\end{tabular}
\end{table*}

A few important conclusions can be drawn from Table\,2. Firstly, the parameter $f$ does not differ
much between the various phases of evolution and cannot be a diagnostic tool to distinguish them.
The small differences in the values of $f$ result rather from slightly various
effective temperatures and luminosities. The result agree with the fact that actually the value of $f$
is determined entirely in subphotospheric layers, thus it mainly depends on the global parameters ($T_{\rm eff},~L/L_{\sun}$).
In other words, the models with the same position in the HR diagram but in various evolutionary stages
would have very similar values of $f$ for a given pulsational mode.
The major effect on the value of the parameter $f$ comes from the efficiency of convective transport in the outer envelope,
measured by the mixing length parameter $\alpha_{\rm MLT}$. As one can see from Table\,2, the models computed for
$\alpha_{\rm MLT}=0.5$ and $\alpha_{\rm MLT}=1.8$ have nearly the same effective temperatures and luminosities
but the values of $f$ for the radial fundamental mode differ enormously. This strong sensitivity is well known
and can be used to derive constraints on $\alpha_{\rm MLT}$ in case of $\delta$ Sct and SX Phe pulsators
if the empirical values of $f$ are determinable from multicolour photometric variability \citep{JDD2003, JDD2005, JDD2007, JDD2020}.
The next finding is that the initial hydrogen abundance $X_0$ and metallicity $Z$ do not affect significantly the values
of $f$ at roughly the same effective temperature and luminosity.
As for the mode instability, in each case the radial fundamental mode is excited. The effect of $\alpha_{\rm MLT}$ on the mode instability
is noticeable and, in this case, the value of $\eta$ is greater for higher values of $\alpha_{\rm MLT}$, i.e., for more efficient convection.

The last conclusion is that having available observables of $\delta$ Scuti, we cannot determine its evolutionary stage;
it can be either the MS or post-MS star. To unravel the evolutionary stage in such case as $\delta$ Scuti, i.e., in the vicinity of TAMS,
independent constraints on the mass and/or radius are necessary.  Such constraints could be reached, e.g., by fitting another
pulsational frequency with a plausible mode identification.
Depending on the evolutionary stage, the age of $\delta$ Scuti is between about 0.82\,Gyr for the main sequence case and about 1.1\,Gyr for the post-main sequence case.

The identification of the radial order $n$ for dipole or quadruple modes  depends strongly on the evolutionary stage.
In the case of the frequency $\nu_2=5.35128$ d$^{-1}$, identified most probably as the dipole mode,
it would be the mode $g_3$ or $g_4$ in the MS or OC model, respectively.
If one considers the post-MS models, then $\nu_2$ would correspond approximately to $g_{18} ... g_{22}$.
To give another example, let us consider the next highest-amplitude frequency in the APT data, i.e., $\nu_5=8.59641$\,d$^{-1}$, as the dipole mode.
In the MS and OC models it will be $p_1$ or $g_1$ while in the post-MS models it will be $g_{8} ... g_{11}$.

In all cases, all these modes have mixed character with the input of the kinetic energy
in the gravity propagation zone, $E_{k,g}$, to the total kinetic energy, $E_k$, up to 90\%.
In Table\,3, we give the examples of the radial order identification for the models from Table\,2 with $Z=0.020$ and $Z=0.028$
at $X_0=0.70$, $\alpha_{\rm ov}=0.25$ and $\alpha_{\rm MLT}=0.5$.  For some modes,
the theoretical frequencies differ quite significantly from the observed values (e.g., in the MS model with $Z=0.02$ and the  OC model with Z=0.028), but one has to remember that the azimuthal order $m$ for $\nu_2$ and $\nu_3$ is unknown.
\begin{table}
\centering
\caption{An assignment of the radial order $n$ to the frequencies $\nu_2$ and $\nu_3$, assuming they are the $\ell=1$ modes,
for the two OPAL models from Table\,2 with the metallicity $Z=0.020$ and $Z=0.028$. The other parameters are: $X_0=0.7$,
$\alpha_{\rm ov}=0.25$ and $\alpha_{\rm MLT}=0.5$. The following columns contain: the evolutionary phase,
mass, the symbol of the observed frequency, the values of the closest theoretical frequency in the model,
mode type, the ratio of $E_{k,g}$ to $E_k$ and the instability parameter $\eta$.}
\begin{tabular}{cccccccccccc}
\hline
\multicolumn{4}{c}{$Z=0.020$} \\
\hline
       &              & observed  &    model  &  mode   &               &        \\
 phase & $M$          & frequency & frequency & type & $E_{k,g}/E_k$ & $\eta$ \\
       & [$M_{\sun}$] &          & [d$^{-1}$] &     &               &        \\
\hline
    MS &     2.25     &   $\nu_2$ &   5.6732  & $g_3$ &      0.83     &  0.081\\
       &              &   $\nu_5$ &   8.6384  & $p_1$ &      0.06     &  0.120\\
\hline
    OC &     2.20     &   $\nu_2$ &   5.3916  & $g_4$ &      0.65     &  0.085\\
       &              &   $\nu_5$ &   8.6812  & $p_1$ &      0.11     &  0.124\\
\hline
post &    2.10     &   $\nu_2$ &   5.3397  & $g_{22}$ &   0.56     &  0.057\\
 -MS      &              &   $\nu_5$ &   8.6754  & $g_{11}$ &   0.56     &  0.119\\
\hline
\multicolumn{4}{c}{$Z=0.028$} \\
\hline
       &              & observed  &    model  &     &               &        \\
 phase & $M$          & frequency & frequency & $n$ & $E_{k,g}/E_k$ & $\eta$ \\
       & [$M_{\sun}$] &          & [d$^{-1}$] &     &               &        \\
\hline
    MS &     2.40     &   $\nu_2$ &  5.2644    & $g_3$  &      0.19     &  0.072\\
       &              &   $\nu_5$ &  8.4742    & $p_1$  &      0.32     &  0.115\\
\hline
    OC &     2.30     &   $\nu_2$ &  5.8684    & $g_4$  &      0.89     &  0.061\\
       &              &   $\nu_5$ &  8.5567    & $g_1$  &      0.34     &  0.110\\
\hline
post &    2.27     &   $\nu_2$ &  5.3362    & $g_{18}$ &    0.71     &  0.047\\
-MS  &              &   $\nu_5$ &  8.7541    & $g_8$   &     0.41     &  0.114\\
\hline
\end{tabular}
\end{table}



\section{Constraints on outer-layer convection }

As have been already mentioned, the method of mode identification described in Sect.\,4 provides, besides the mode degree $\ell$, semi-empirical values of two complex quantities: the parameter $f$ and the intrinsic mode amplitude $\varepsilon$ multiplied by $Y_\ell^m(i,0)$.
As a reminder, the parameter $f$ is the amplitude of the radiative flux variations at the level of the photosphere and $\varepsilon$ gives
the relative radius variation. These two parameters are semi-empirical because their values
depend on the model atmospheres, in particular on the parameters of the metallicity [m/H] and microturbulent velocity $\xi_t$.
The quantity $\tilde\varepsilon= \varepsilon Y^m_{\ell}(i,0)$ has been already used to make a choice between the three mode degrees
for the frequency $\nu_2=5.35128$\,d$^{-1}$. We obtained also that the dominant mode causes a change of the photospheric radius
by 0.3-0.9\,\%.

Unlike $\varepsilon$, the empirical values of $f$ can be directly compared with the results of linear nonadiabatic
computations of stellar pulsations provided that the mode degree $\ell$ is uniquely identified.
We started pulsational modelling with the OPAL opacity tables assuming five values of the mixing length parameter
$\alpha_{\rm MLT}=0.0,~0.5,~1.0,~1.8,~2.5$.

In Fig.\,11, we show comparisons of the empirical and theoretical values of $f$ for the two models fitting frequency $\nu_1$ as the radial fundamental mode. The real and imaginary part of $f$ is denoted by $f_R$ and $f_I$, respectively.
The top panel shows the theoretical values of $f$ for the model with $Z=0.02$, $M=2.25~M_{\sun}$, $\log T_{\rm eff}=3.8408$ and $\log L/L_{\sun}= 1.571$.
The corresponding empirical values are for an atmospheric metallicity $\mbox{[m/H]}=0.2, 0.3$ and a microturbulent velocity $\xi_t=2,~4~\kms$.
In the bottom panel a comparison is made for the theoretical model with the parameters: $Z=0.028$, $M=2.40~M_{\sun}$,
$\log T_{\rm eff}=3.8383$, $\log L/L_{\sun}= 1.578$, and the empirical values of $f$ were determined for $\mbox{[m/H]}=0.3, 0.5$
and $\xi_t=2,~4~\kms$. Both these models had an initial hydrogen abundance $X_0=0.70$ and overshooting from the convective
core $\alpha_{\rm ov}=0.25$. We look for models (orange dots) which are within the errors of the empirical values (black dots).
\begin{figure}
\includegraphics[width=\columnwidth,clip]{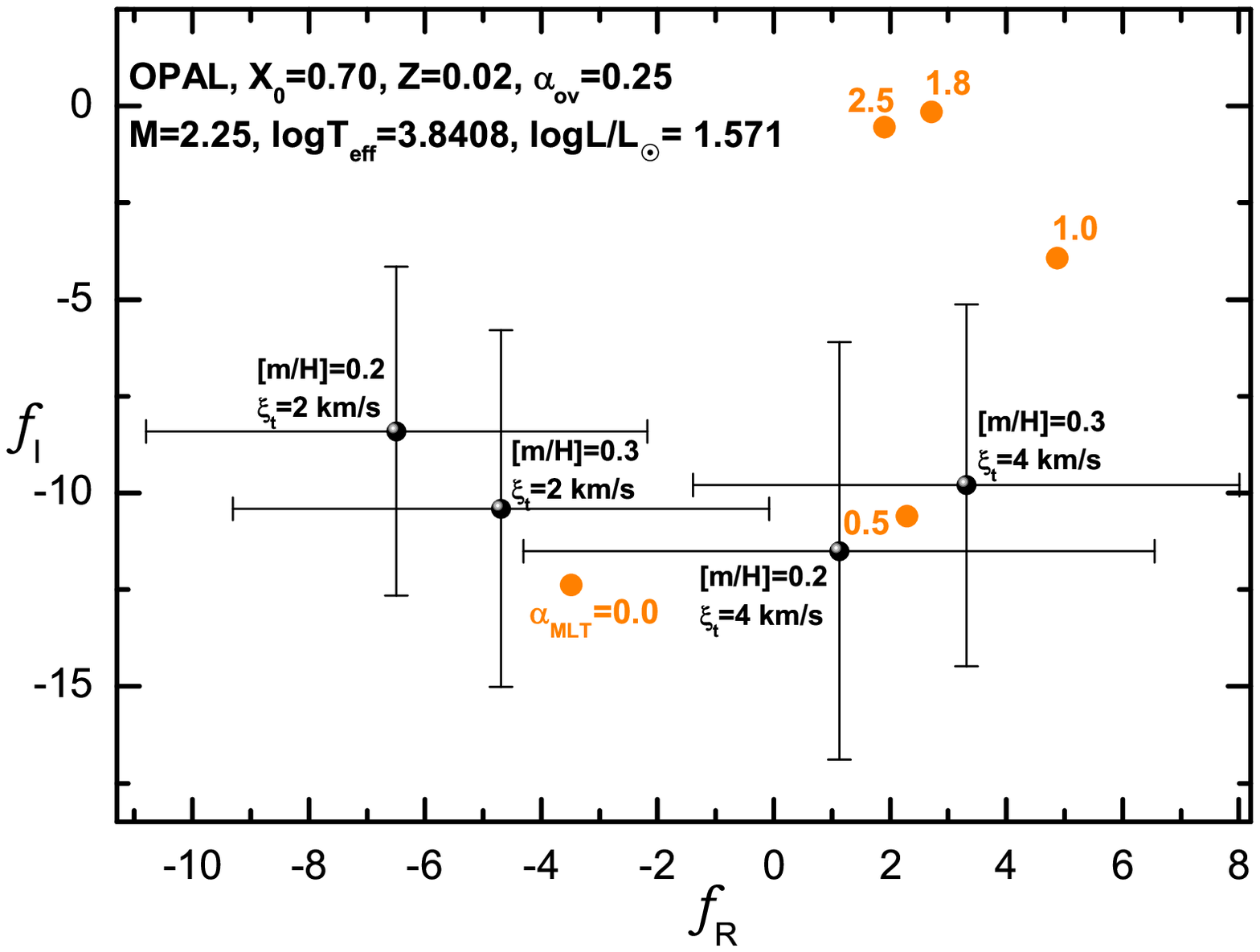}
\includegraphics[width=\columnwidth,clip]{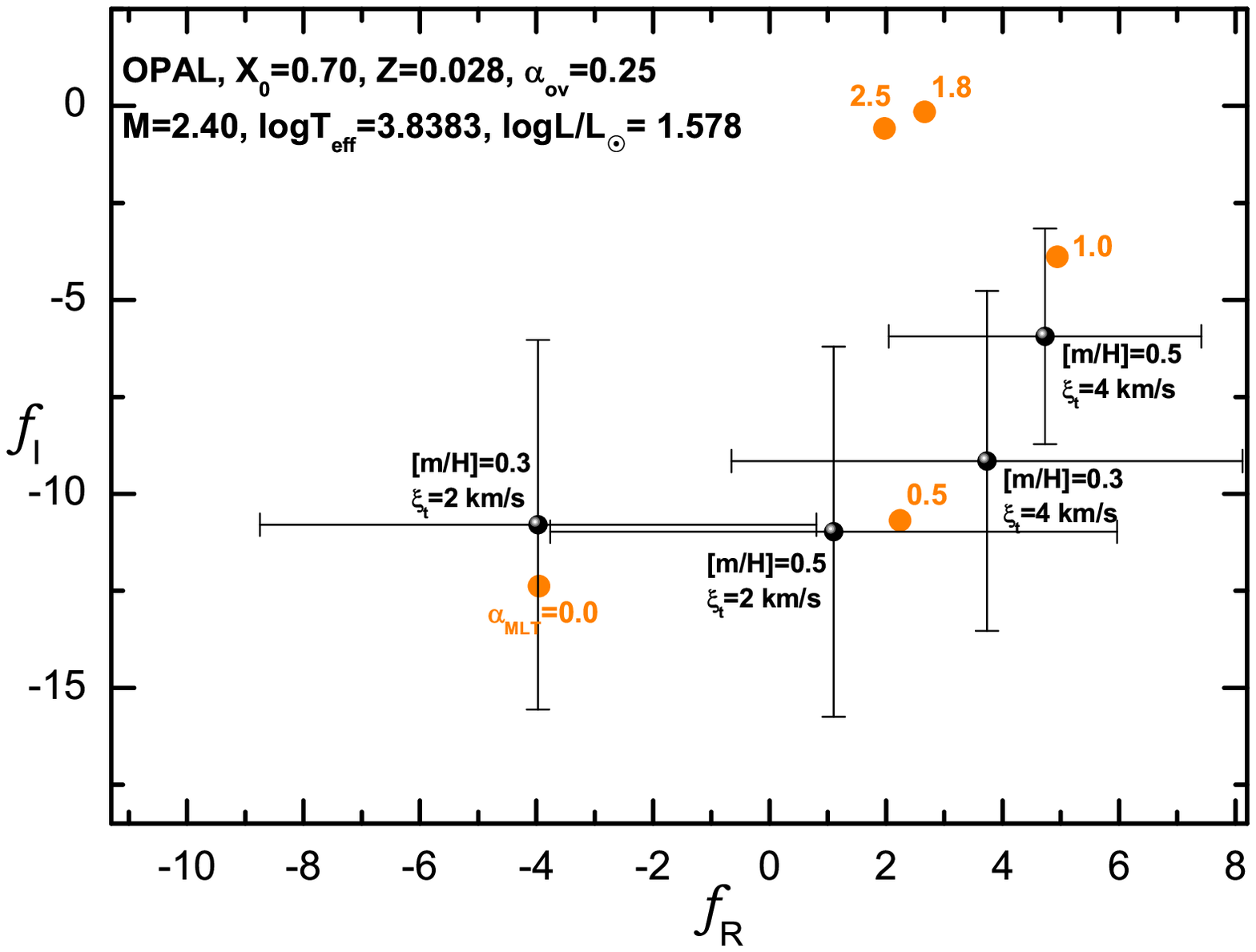}
	\caption{A comparison of the theoretical and empirical values of the nonadiabatic parameter $f$ for the two OPAL models that reproduce
the dominant frequency as the radial fundamental mode. The theoretical values of $f$ were computed for five values of the mixing length
parameter $\alpha_{\rm MLT}=0.0,~0.5,~1.0,~1.8$ and 2.5 (orange dots). The top panel shows the results for $Z=0.020$
and the bottom panel for $Z=0.028$.
The parameters of the models are given in the text.
The empirical values are shown for the atmospheric metallicity $\mbox{[m/H]}=0.2, 0.3$ (the top panel)
and $\mbox{[m/H]}=0.3, 0.5$ (the bottom panel). Two values of the microturbulent velocity $\xi_t$ were adopted in each case.}
\label{fig11}
\end{figure}

As we can see, despite the large errors in the empirical values of $f$, we can conclude that the efficiency of convective transport
in the envelope of $\delta$ Scuti can be described by the mixing length parameter $\alpha_{\rm MLT}$ less than about 1.0.
This conclusion is independent of the input parameters in evolutionary computations, i.e., $Z$, $X_0$ and $\alpha_{\rm ov}$,
as well as  independent of  $T_{\rm eff}$ and $\log L$ in the observed ranges of $\delta$ Scuti.
On the other hand, one can see a strong effect of the atmospheric parameters, i.e., [m/H] and $\xi_t$,
on the empirical values of $f$. This effect is comparable to the effect of $\alpha_{\rm MLT}$ on the theoretical values of $f$.
In case of the models computed for $Z=0.020$ (the top panel of Fig.\,11), we obtain approximately:
\begin{itemize}
\item $\alpha_{\rm MLT}\lesssim 0.5$ for  $\xi_t=2~\kms$,
\item $\alpha_{\rm MLT}\lesssim 1.0$ for $\xi_t=4~\kms$.
\end{itemize}
In the case of the models with $Z=0.028$, we have:
\begin{itemize}
\item $\alpha_{\rm MLT}\lesssim 0.5$ for $\mbox{([m/H]},~\xi_t)=(0.3,~2~\kms)$,
\item $\alpha_{\rm MLT}\lesssim 1.0$ for $\mbox{([m/H]},\xi_t)=(0.3,4\kms)\cup(0.5,2\kms)$,
\item $\alpha_{\rm MLT}\in(0.5,~1.0)$ for $\mbox{([m/H]},~\xi_t)=(0.5, 4~\kms)$.
\end{itemize}

Because for the pair ([m/H], $\xi_t$)=(0.5, $4~\kms$) we obtained the most unique identification of $\ell$ in Sect.\,4,
we may infer, cautiously, that the most likely range of $\alpha_{\rm MLT}$ is (0.5,\,1.0).
Nonetheless, regardless of the evolutionary and atmospheric parameters, we can conclude that the mixing length parameter $\alpha_{\rm MLT}$
in the outer layers of $\delta$ Scuti is less than about 1.0.

Now the question is how this constraint on  $\alpha_{\rm MLT}$ depends on the adopted opacity data. To study this effect
we computed pulsational models that reproduce the dominant frequency of $\delta$ Scuti as the radial fundamental mode 
adopting the OP tables \citep{Seaton2005} and OPLIB data \citep{Colgan2016}.
In Fig.\,12, we show the values of $f$ for all three sources of the opacity data for $X_0=0.70$, $Z=0.028$ and $\alpha_{\rm ov}=0.25$.
The stellar parameters of these models slightly change for various values of $\alpha_{\rm MLT}$ but
the differences are in the fourth decimal place and the average stellar parameters are as follows:
\begin{itemize}
\item {\bf OPAL}:~~$M=2.40M_{\sun}$, $\log T_{\rm eff}=3.8383$, $\log L/L_{\sun}= 1.578$,
\item {\bf OP}:~~~~~~~$M=2.40M_{\sun}$, $\log T_{\rm eff}=3.8379$, $\log L/L_{\sun}= 1.576$,
\item {\bf OPLIB}: $M=2.42M_{\sun}$, $\log T_{\rm eff}=3.8378$, $\log L/L_{\sun}= 1.580$.
\end{itemize}
The parameters of the OPAL model are the same as in the lower panel of Fig.\,11 but for a direct comparison we repeated them.
The empirical values of $f$ plotted in Fig.\,12 are also the same as in the lower panel of Fig.\,11
and the differences in their values for the above mentioned parameters of the OP and OPLIB models are 10\% at most.
Therefore, for the sake of clarity, we did not plot all empirical values of $f$.
\begin{figure}
\includegraphics[width=\columnwidth,clip]{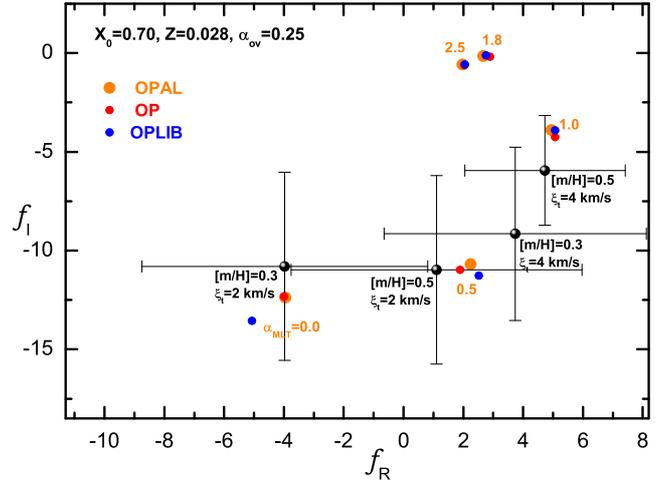}
\caption{The same as in the bottom panel of Fig.\,11, but the result for the OP and OPLIB models with very close stellar parameters were added.}
\label{fig12}
\end{figure}

The consistency of the theoretical values of $f$ in the models of $\delta$ Scuti for the different opacity tables is striking.
This result is opposite to what we obtained for B-type pulsators in our several studies  \citep[e.g.,][]{JDD2005b, JDD2017, Walczak2019}.
Thus, independently of the adopted opacities the constraints on the values of $\alpha_{\rm MLT}$ remain unchanged.

Finally, we examined the effect of the chemical mixture as the atmospheric abundances of $\delta$ Scuti are anomalous.
Such a check is important because  the values of $f$ are determined in the subphotospheric layers.
In Fig.\,13, we compare the values of $f$  of the OPAL models computed with the AGSS09 mixture
and the $\delta$ Sct mixture  as determined by \citet{Yushchenko2005}.
As in Fig.\,12,  we adopted $X_0=0.70$, $Z=0.028$ and $\alpha_{\rm ov}=0.25$
The stellar parameters of the OPAL model obtained with the $\delta$ Sct mixture,
that fit the frequency of the dominant mode, are:
\begin{itemize}
	\item {\bf OPAL($\delta$ Sct)}:~~$M=2.38M_{\sun}$, $\log T_{\rm eff}=3.8385$, $\log L/L_{\sun}= 1.577$
\end{itemize}
 As one can see, the effect of the chemical mixture  is even more negligible than when using different opacity tables.

\begin{figure}
\includegraphics[width=\columnwidth,clip]{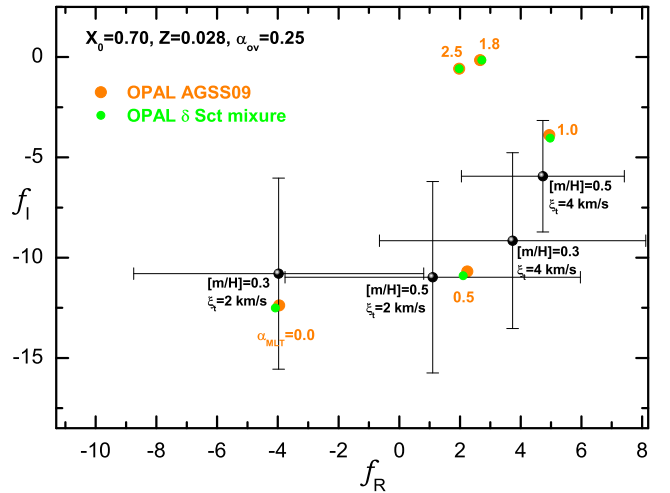}
\caption{The same as in the bottom panel of Fig.\,11, but the result for the OPAL models computed with  the $\delta$ Scuti mixture adopted from \citet{Yushchenko2005} were added.}
\label{fig13}
\end{figure}

The above results validate the obtained limits on the efficiency of convection in the $\delta$ Scuti envelope.
Moreover, almost in all models the radial fundamental mode is excited. The exceptions are the OPAL and OPLIB models
with a mixing length parameter $\alpha_{\rm MLT}=2.5$.

\section{Conclusions}
The aim of this work was to present an extended analysis of the pulsating star $\delta$ Scuti,
the class prototype that is far too little observed and studied.
We analysed new multi-colour time-series photometry of the star in order to extract
the pulsational frequencies and, then, to identify their mode degrees.
Applying Fourier analysis to the seven-year SMEI data we extracted 18 significant frequencies with 14 being independent.
No low-frequency peaks were found that could correspond to high-order gravity modes.
In the combined $uvy$ APT data, we detected only 10 out of 18 SMEI peaks and no other signals.
Eight APT frequencies are independent and correspond to the highest-amplitude SMEI peaks but $\nu_3=8.376999$\,d$^{-1}$.
A re-analysis of the data from \citet{Templeton1997} revealed 9 significant frequencies with one undetected in the SMEI and APT observations.
The third highest-amplitude frequency $\nu_3=8.376999$\,d$^{-1}$ detected in SMEI was also not present in Templeton et al.'s data.
The lack of frequency $\nu_3$ in the APT and Templeton et al. data proves a huge change in the amplitude of this pulsation mode.

Having the amplitudes and phases in three passbands, i.e., the Str\"omgren $uvy$, we made an attempt to identify
the harmonic degree $\ell$ of the detected modes in $\delta$ Scuti, using the method of \citet{JDD2003}.
To this end we used Vienna model atmospheres that take into account turbulent convection. We obtained that the dominant frequency is unequivocally
a radial mode as already has been suggested by, e.g., \citet{Balona1981}, \citet{Cugier1993} and \citet{Templeton1997}.

$\delta$ Scuti is quite abundant in metals with a measured value of metallicity $\mbox{[m/H]}\in(0.18,0.38)$.
It is also considered as a chemical peculiar star with
an excess of such elements as  aluminium, iron, nickel and manganese. This is probably a reason
why the goodness of the fit of the calculated amplitudes and phases to the observed values are so sensitive to the adopted metallicity [m/H] and microturbulent velocity $\xi_t$. This sensitivity is even stronger for the lower-amplitude frequencies.
The most homogenous fit  in the whole observed range of $T_{\rm eff}$
was obtained for [m/H]=+0.5 and $\xi_t=4~\kms$. In other words, for these atmospheric parameters
the discriminant $\chi^2$ has a similar dependence on the effective temperature and mode degree $\ell$.
We determined that frequency $\nu_2$ is most probably an $\ell=1$ mode and the other six frequencies have far less unique identifications of $\ell$.
The identification of $\ell=1$ for $\nu_2$ was also verified by comparing the estimated amplitude of the radial velocity change
with the observed value as derived by \citet{Balona1981}. This assessment has been done by
adopting the empirical values of ${\tilde\varepsilon}\equiv \varepsilon Y^m_{\ell}(i,0)$ as determined from our method.
Moreover, for the dominant frequency $\nu_1$, which is a radial mode, we could estimate the intrinsic amplitude itself, i.e., $\varepsilon$.
The result is that the changes of the radius caused by the dominant mode are between 0.3\% and 0.9\%. 

In the next step, we constructed the pulsational models that reproduce the dominant frequency as the radial fundamental
or first overtone modes. It appeared that in the allowed range of $(T_{\rm eff},~L/L_{\sun})$, the frequency $\nu_1$ can be associated
only with the fundamental mode, which confirms Templeton et al.'s result. The pulsational models with $\nu_1$ as the first overtone
are far outside the error box and all are in the post-main sequence phase. In the case of the radial fundamental mode,
three phases of evolution are possible: main sequence, overall contraction and post-main sequence stage.
However, for the star to be on the main sequence, the convective overshooting parameter, $\alpha_{\rm ov}$ ,
must be at least 0.25 for the metallicity greater than $Z=0.019$.

Unfortunately, we do not have any diagnostic tool that can distinguish what the actual evolution phase of $\delta$ Scuti is.
For example, it turned out that the value of the nonadiabatic parameter $f$
is determined by the effective temperature and the luminosity, i.e.,  by the position of the star in the HR diagram.
Moreover, there are no more frequencies with a firm identification of $(\ell,~m)$, that could better constrain our seismic models.
Thus, we are left with the open question of whether this prototype is a main-sequence or post-main sequence star.
In the considered range of the initial hydrogen abundance $X_0$ and metallicity $Z$,
our estimate of the age of $\delta$ Scuti is about 0.82-0.89\,Gyr for the main sequence phases,
about 0.92-1.02\,Gyr for the overall contraction and about 1.02-1.09\,Gyr for the post-main sequence phase.

One of the main results of this work are constraints on the efficiency of convective transport in the envelope of $\delta$ Scuti,
on the assumption of the frozen convection approximation.
This has been done by comparing the empirical and theoretical values of the parameter $f$ for the dominant radial mode.
Our results testify in favour of rather low to moderately efficient convection in the outer layers, described
by a mixing length parameter $\alpha_{\rm MLT}$ less than 1.0, which is significantly lower than the solar value $\alpha_{\rm MLT}$=1.8.
These constraints on $\alpha_{\rm MLT}$ are independent
of the commonly-used opacity data as the same result was obtained with all three opacity tables: OPAL, OP and OPLIB.
This weak sensitivity of $f$ on the adopted opacity data is contrary to what we have learned from our studies of B-type pulsators
\citep[e.g.,][]{JDD2005b, JDD2017, Walczak2019}. Moreover, the effect of the chemical mixture is completely negligible.

Unambiguous identification of the mode degree $\ell$ for more frequencies of $\delta$ Scuti could also help to further refine
the parameter  $\alpha_{\rm MLT}$. That requires more multicolour photometric observations. Besides, a simultaneous spectroscopic campaign
would allow to detect more frequencies, to improve mode identification, including determination of the azimuthal order $m$,
and to constrain the atmospheric metallicity [m/H] and microturbulent velocity $\xi_t$ for a better comparison
of the theoretical and empirical values of $f$.

Finally, experience with space photometry of other $\delta$ Scuti stars suggests that low-frequency oscillations would likely as well be detected in $\delta$ Scuti itself.
Then, a more comprehensive seismic study could be carried out to explain the excitation of high-order gravity modes
and possibly to achieve constraints on the mean opacity profile.

\section*{Acknowledgements}
The work was financially supported by the Polish NCN grant 2018/29/B/ST9/02803.
GH thanks the Polish National Center for Science (NCN) for support through grant 2015/18/A/ST9/00578.
APi acknowledges support from the NCN grant no. 2016/21/B/ST9/01126.
We are grateful to Matthew Templeton for sharing his published data and to Javier Pascual Granado for obtaining additional photometry of the star.
This work has made use of data from the European Space Agency (ESA) mission
{\it Gaia} (\url{https://www.cosmos.esa.int/gaia}), processed by the {\it Gaia}
Data Processing and Analysis Consortium (DPAC,
\url{https://www.cosmos.esa.int/web/gaia/dpac/consortium}). Funding for the DPAC
has been provided by national institutions, in particular the institutions
participating in the {\it Gaia} Multilateral Agreement.
\section*{Data Availability}
The APT observations as well as theoretical computations will be shared on reasonable request to the corresponding author.



\bibliographystyle{mnras}
\bibliography{JDD_biblio} 


\bsp	
\label{lastpage}
\end{document}